\documentclass[12pt]{article}

\input prepictex.tex
\input pictex.tex
\input postpictex.tex

\begin{document}
\begin{center}
\vspace{.8cm}
\LARGE
{\bf Einstein's Apple}
\footnote[1]{The text of this paper has been incorporated
as various chapters in a book of the same name, that is,
{\it Einstein's Apple}; which is available for download as
a PDF on Schucking's NYU Physics Department webpage
http://physics.as.nyu.edu/object/EngelbertSchucking.html}
\\
\Large
{\bf His First Principle of Equivalence}\\

\vspace{.5cm}
{\bf \today} \\
\vspace{.5cm}

\large{Engelbert Schucking} \\
\vspace{.15cm}
\normalsize
Physics Department \\
New York University \\
4 Washington Place \\
New York, NY 10003 \\
elschucking@msn.com \\
\vspace{.5cm}
\large{Eugene J. Surowitz} \\
\vspace{.15cm}
\normalsize
Visiting Scholar \\
New York University \\
New York, NY 10003 \\
surow@attglobal.net \\
\vspace{.8cm}

\end{center}

\begin{abstract}
After a historical discussion of Einstein's 1907
principle of equivalence, a homogeneous gravitational field
in Minkowski spacetime is constructed.
It is pointed out that the reference frames in gravitational
theory can be understood as spaces with a flat connection
and torsion defined through teleparallelism.
This kind of torsion was first introduced by Einstein in 1928.
The concept of torsion is discussed through simple examples
and some historical observations.
\end{abstract}

\eject
\section{The Principle of Equivalence}
\setcounter{equation}{0}

In a speech given in Kyoto, Japan, on December 14, 1922
Albert Einstein remembered:
\begin{quote}
\lq\lq I was sitting on a chair in my patent office in Bern.
Suddenly a thought struck me:
If a man falls freely, he would not feel his weight.
I was taken aback.
The simple thought experiment made a deep impression on me.
It was what led me to the theory of gravity.\rq\rq
\end{quote}
This epiphany,
that he once termed  \lq\lq {\it der gl\"ucklichste Gedanke meines Lebens}\,\rq\rq\
(the happiest thought of my life),
was an unusual vision in 1907.
Einstein's apple was not envisioned after watching the antics
of orbiting astronauts on television, sky-diving clubs did not yet exist,
and platform diving was not yet a sports category of the freshly
revived Olympic Games.
How could this thought have struck him?
Had he just been dealing with patent applications covering
the safety of elevators?

Three years earlier, in 1904, the Otis Elevator Company installed in Chicago, Illinois,
the first gearless traction electric elevator apparatus,
that was of the direct drive type, known as the \lq\lq 1:1 elevator\rq\rq.
This first modern electric elevator made its way to Europe where,
on Z\"urich's Bahnhofstrasse and elsewhere in Switzerland,
buildings went up that needed elevators.
It would have been natural for Director Friedrich Haller at the Swiss Patent Office
in Bern to put applications involving electro-mechanical machinery
on the desk of Einstein, his expert, second class, with expertise
in electromagnetism.
It would not be surprising to find Einstein's signature approving
(or disapproving) a patent application for elevator's in 1907.

However, only one patent application with Einstein's comments from his
years as a Swiss patent examiner has survived.
It does not concern elevators \cite{Fluckinger}.
The comments of the patent examiners in Haller's files have not
been preserved by the Swiss bureaucracy.
Einstein's expert opinions on patents were destroyed eighteen years
after the files were closed.
We'll probably never know how Einstein got his inspiration.

In his review
\lq\lq The Relativity Principle and the Conclusions drawn from it\rq\rq
\cite{Einstein1907},
Einstein formulated his principle of equivalence for the first time.
He wrote:
\begin{quote}
\lq\lq We consider two systems $\Sigma_{1}$ and $\Sigma_{2}$ in motion.
Let $\Sigma_{1}$ be accelerated in the direction of its $X$-axis, and
let $\gamma$ be the (temporally constant) magnitude of that acceleration.
$\Sigma_{2}$ shall be at rest, but it shall be located in a homogeneous
gravitational field that imparts to all objects an acceleration $- \gamma$
in the direction of the $X$-axis.\rq\rq
\end{quote}
The next sentence contains the principle of equivalence:
\begin{quote}
\lq\lq As far as we know, the physical laws with respect to $\Sigma_{1}$
do not differ from those with respect to $\Sigma_{2}$;
this is based on the fact that all bodies are equally accelerated
in the gravitational field.\rq\rq
\end{quote}
It was this last fact that had inspired Einstein and prompted
Sir Hermann Bondi, Master of Churchill College in Cambridge, England,
to the observation:
\begin{quote}
\lq\lq If a bird watching physicist falls off a cliff,
he doesn't worry about his binoculars, they fall with him.\rq\rq
\end{quote}
Although nowhere stated in the \lq\lq{\it Principia}\,\rq\rq,
one may assume that Isaac Newton was already familiar with
the principle of equivalence.
In Proposition 6 of Book 3 of his Principia\cite{Newton}, Newton describes
his precise pendulum experiments with
\lq\lq gold, silver, lead, glass, sand, common salt, wood,
water and wheat\rq\rq\
testing equality---as we would say now---of inertial and passive
gravitational mass.
His treatment in Proposition 26 of Book 3 in the Principia
dealing with the perturbation by the Sun of the lunar orbit
around the Earth, discovered by Tycho Brahe and called the
\lq\lq Variation,\rq\rq\ leaves no doubt that Newton knew
how to transform away a homogeneous gravitational field.

The apparent enigmatic equality of inertial and passive gravitational
mass was also still a prize question at the beginning of the
twentieth century.
The Academy of Sciences in G\"ottingen, Germany, had offered
the Beneke Prize in 1906 for proving this equality by experiment
and theory.
The Baron Roland E\"otv\"os won three-fourths of this prize
(3,400 of 4,500 Marks); only three-fourths, because he had only done the
experiments and had not attempted a theoretical explanation \cite{Runge}.

The principle of equivalence was not new in Newton's theory
of gravitation.
New was Einstein's extension to all of physics.
He wrote:
\begin{quote}
\lq\lq At our present state of experience we have thus no reason to assume
that the systems $\Sigma_{1}$ and $\Sigma_{2}$ differ from each other
in any respect, and in the discussion that follows,
we shall therefore assume the complete physical equivalence of
a gravitational field and a corresponding acceleration of the
reference system.\rq\rq
\end{quote}
Before we go on with the discussion, we have to say more about
the reference system.

\section{The Reference System}
\setcounter{equation}{0}

Einstein's reference system was based on identical clocks and rigid bodies.
Through the work of Gustav Herglotz, Max Born, and Max von Laue,
it was soon realized that rigid bodies do not form suitable
reference systems.
The ideas of Hermann Minkowski and Henri Poincar\'e allow us to describe
the reference systems of special relativity more clearly.
The Minkowski spacetime with a metric given by the line element 
\begin{equation}
ds^{2} \, =  \, \eta_{\mu\nu} \, dx^{\mu} dx^{\nu} \, ; \qquad \mu, \nu \in 0,1,2,3 \, ,
\end{equation}
with
\begin{equation}
\eta_{\mu\nu} \, = \, \left(
\begin{array}{cccc}
1 & 0 & 0 & 0 \\ 0 & -1 & 0 & 0 \\ 0 & 0 & -1 & 0 \\ 0 & 0 & 0 & -1
\end{array}
\right)
\end{equation}
has a set of distinguished coordinates $x^{\mu}$.
Under a Poincar\'e transformation
\begin{equation}
x'{}^{\mu}
\, = \,
\Lambda^{\mu}{}_{\nu} x^{\nu} \, + a^{\mu} \, , \qquad a^{\mu}
\, = \, constant
\end{equation}
with constant $ \Lambda^{\mu}{}_{\nu} $ and
\begin{equation}
\Lambda^{\mu}{}_{\nu} \, \eta_{\mu\rho} \,
\Lambda^{\rho}{}_{\sigma} \, = \, \eta_{\nu\sigma}
\end{equation}
the $ x'{}^{\mu} $ again form such a set of distinguished coordinates.
We call them linear orthonormal coordinates.

To keep physics and mathematics clearly defined,
we introduce four constant orthonormal vector fields
$ {\bf e}_{\mu} $ such that
\begin{equation} \label{eq:4ConstantOrtho}
{\bf e}_{\mu} \, \equiv \, \frac{\partial \quad }{\partial \bf x^{\mu}}
\end{equation}
for a set of distinguished coordinates $x^{\mu}$.
The orthonormal vectors along the coordinate axes just introduced
by the operators $ \partial / \partial x^{\nu} $ are subject to 
\begin{equation} \label{eq:RefSysdxddx}
{\mbox {\boldmath $dx$}}^{\mu} 
\left( \frac{\partial \quad}{\partial x^{\nu}} \right)
\, = \, \delta^{\mu}{}_{\nu} \, .
\end{equation}

The invariant characterization of a vector tangent to a manifold as a
directional differentiation operator on the functions living on this manifold
goes back to Sophus Lie.
He used that concept in his \lq\lq {\it Theory of Transformation Groups}\rq\rq,
where vector fields became \lq\lq infinitesimal transformations\rq\rq.
Defining differentials like $ dx^{\mu} $ as the duals of vectors
like $ \partial/\partial x^{\mu} $ emerged from the work of \'Elie Cartan
and its interpretation by Erich Kaehler and Nicholas Bourbaki.

The vectors in
Eq.(\ref{eq:RefSysdxddx})
are all parallel to each other
and defined on all of Minkowski spacetime.
The components $ V^{\nu}(x) $ of any vector field {\bf V}
\begin{equation}
{\bf V} \, = \, V^{\nu}(x) \left( \frac{\partial \quad}{\partial x^{\nu}} \right) \, ,
\end{equation}
also known as the \lq\lq physical\rq\rq\ components, have a metrical meaning,
that is to say, their numerical values are results of physical measurements
in terms of meters or seconds.

If one introduces new coordinates $ y^{\alpha} $ that are non-linear
and/or non-orthonormal functions of the distinguished coordinates $ x^{\nu} $
\begin{equation}
y^{\alpha} \, = \, f^{\alpha}(x^{\nu}) \, ,
\qquad \det \left[ \, \frac{\partial y^{\alpha}}{\partial x^{\nu}} \, \right]
\, \neq \, 0 \, ,
\end{equation}
the basis vector fields $ \partial / \partial y^{\alpha} $ will no longer
be orthonormal and the components $ V^{\alpha}(y) $ of
\begin{equation}
{\bf V} \, = \, V^{\alpha}(y) \left( \frac{\partial \quad}{\partial y^{\alpha}} \right)
\end{equation}
will no longer be physical components.
The metric tensor $ g_{\alpha\beta} $
\begin{equation}
ds^{2} \, = \, g_{\alpha\beta}(y) \,
\frac{\partial y^{\alpha}}{\partial x^{\mu}} \,
\frac{\partial y^{\beta}}{\partial x^{\nu}} \,
dx^{\mu} \, dx^{\nu}
\, = \,
\eta_{\mu\nu} \, dx^{\mu} \, dx^{\nu}
\end{equation}
will no longer be $ \eta_{\alpha\beta} $.

If we take the mathematician's point of view that
coordinates are an arbitrary means for naming events in Minkowski spacetime,
we cannot assign a direct physical meaning
to the components of vector fields given in those coordinates either.
However, this was not a point of view of taken by Einstein originally and
became the root of misunderstandings.

If we want to know the physical components $ V^{\mu}(x) $
in arbitrary coordinates $ y^{\alpha}(x^{\mu}) $,
we only have to remember the chain rule to see that
\begin{equation}
V^{\mu}(x) \, \frac{\partial \quad}{\partial x^{\mu}}
\, = \,
V^{\alpha}(y) \, \frac{\partial \quad}{\partial y^{\alpha}} \,
\, = \,
V^{\mu}(x) \, \frac{\partial y^{\alpha}(x)}{\partial x^{\mu}} \,
\frac{\partial \quad}{\partial y^{\alpha}} \, .
\end{equation}
This gives the relation
\begin{equation}
V^{\alpha}(y)
\, = \,
V^{\mu}(x) \, \frac{\partial y^{\alpha}(x)}{\partial x^{\mu}}
\end{equation}
for the transformation of vector components from one
coordinate system to another.

The use of arbitrary coordinates in Minkowski spacetime is just
a matter of convenience for adapting the coordinates to the
symmetry of a given situation.
We should now stress that
the identification of the constant vector field with the $ {\bf e}_{\mu} $,
as defined in
Eq.(\ref{eq:4ConstantOrtho}),
is invariant under the Poincar\'e group.
The translations evidently do nothing and
constant Lorentz transformations give
\begin{equation}
{\bf e}_{\nu} \, = \, {\bf e'}_{\mu} \, \Lambda^{\mu}{}_{\nu}
\qquad \Longleftrightarrow \qquad
\frac{\partial \quad }{\partial x^{\nu}}
\, = \,
\frac{\partial \quad }{\partial x'{}^{\mu}} \, \Lambda^{\mu}{}_{\nu} \, .
\end{equation}
We call the set of four constant orthonormal vectors, $ {\bf e}_{\nu} $,
in each event of Minkowski spacetime a \lq\lq frame\rq\rq.

In the nineteenth century physicists were already using more general frames
in 3-dimens\-ional space.
For instance,
in problems with axial or spherical symmetry
it was useful to let the frame vectors $ {\bf i}, {\bf j}, {\bf k} $ be tangent to the
orthogonal lines of constant coordinate pairs.
In this way one could define, for example,
the radial component of an electrical field strength.
With the advent of Minkowski's spacetime in 1908,
such adaptable orthonormal frames also appeared in four dimensions.

The example of Born motion is instructive.

\section{Born Motion} \label{sect:BM}
\setcounter{equation}{0}

Born's motion stands in analogy to an observer's motion on the circle
\begin{equation}
x^{2} \, + \, y^{2} \, = \, \rho^{2}
\end{equation}
of radius $ \rho $ in the $x$-$y$--plane.
If the motion is at constant speed along the circle,
an observer experiences a constant acceleration
in the direction opposite from the center of motion.
In Born motion we require an observer to feel constant acceleration
in Minkowski space-time.

We introduce into the $x$-$t$--plane (with $ c = 1 $) \lq\lq polar\rq\rq\ coordinates
\begin{equation} \label{eq:BMpolarDef}
t \, = \, \rho \ \sinh \, \tau \, ,
\qquad
x \, = \, \rho \ \cosh \, \tau \, ,
\end{equation}
and they yield the line element in the form
\begin{equation} \label{eq:BMpolar}
ds^{2} \, = \, dt^{2} \, - \, dx^{2} \, = \, \rho^{2} \, d\tau^{2} \, - \, d\rho^{2} \, ,
\end{equation}
since the differentials of
Eq.(\ref{eq:BMpolarDef})
are related by
\begin{equation}
dt \, = \, d\rho \ \sinh \, \tau \, + \, \rho \ \cosh \, \tau \ d\tau \, ,
\qquad
dx \, = \, d\rho \ \cosh \, \tau \, + \, \rho \ \sinh \, \tau \ d\tau \, .
\end{equation}
The vector form of the line element is
\begin{eqnarray}
\frac{\partial \ }{\partial s} \otimes \frac{\partial \ }{\partial s}
\, & = & \,
\frac{1}{\rho^{2}} \,
\frac{\partial \ }{\partial \tau} \otimes \frac{\partial \ }{\partial \tau}
\, + \,
\frac{\partial \ }{\partial \rho} \otimes \frac{\partial \ }{\partial \rho}
\\ \nonumber \\
\, & = & \,
{\bf e}_{0} \otimes \, {\bf e}_{0}
\, + \, {\bf e}_{1} \otimes \, {\bf e}_{1} \, ,
\end{eqnarray}
where the vector fields
\begin{equation} \label{eq:BMvectors}
{\bf e}_{0} \, \equiv \, \frac{1}{\rho} \, \frac{\partial \ }{\partial \tau} \, ,
\qquad
{\bf e}_{1} \, \equiv \, \frac{\partial \ }{\partial \rho}
\end{equation}
are orthonormal.
The coordinate lines $ \rho = constant $ are the hyperbolae
\begin{equation} \label{BornHyperbolae}
x^{2} \, - \, t^{2} \, = \, \rho^{2}
\end{equation}
of Born motion.
Its time-like worldlines, with proper time $s$ as parameter,
\begin{equation}
\tau \, = \, \frac{s}{\rho} \, , \qquad \rho \, = \, constant \, ,
\end{equation}
have the unit tangent vector
\begin{equation}
\left( \frac{dt}{ds} \, , \frac{dx}{ds} \right)
\, = \,
\left( \cosh \frac{s}{\rho} \, , \, \sinh \frac{s}{\rho} \right)
\, \equiv \,
{\bf e}_{0} \, .
\end{equation}
The acceleration vector is given by
\begin{equation}
\left( \frac{d^{2}t}{ds^{2}} \, , \frac{d^{2}x}{ds^{2}} \right)
\, = \,
\frac{1}{\rho} \left( \sinh \frac{s}{\rho} \, , \, \cosh \frac{s}{\rho} \right)
\, \equiv \,
\frac{1}{\rho} \ {\bf e}_{1} \, .
\end{equation}
This shows that $ 1 / \rho $ is the amount of acceleration
and the hyperbolae Eq.(\ref{BornHyperbolae}) are the worldlines of constant acceleration.

For Born motion,
acceleration has become the curvature of the circular motion and the hyperbolic functions
turn into trigonometric ones.

It is easy to construct a picture of the vector fields
$ {\bf e}_{0} $ and $ {\bf e}_{1} $ of Eq.(\ref{eq:BMvectors}) that show,
in a region of Minkowski spacetime,
the local four-velocity and four acceleration.
[{\bf See Figure~0}.]
$$
\hskip 1.5 in
{\beginpicture
    \setcoordinatesystem units <1.0in,1.0in> point at 0 0

    \arrow <6pt> [.1,.4] from  -2.5    0.0   to   2.5   0.0  
    \arrow <6pt> [.1,.4] from   0.0   -0.5   to   0.0   3.0  

    \put {$ {\bf \tau}$} at -0.15   2.95  
    \put {$ {\bf \rho}$} at  2.45  -0.15  

    \put { {\large e}$_0$
           $ { \displaystyle  \, = \,
               \frac{1}{\rho} \frac{\partial \ }{\partial \bf \tau} }
           $ } at -0.60   2.50  

    \put { {\large e}$_1$
           $ {\displaystyle  \, = \,
             \frac{\partial \ }{\partial \bf \rho} }
          $} at  1.90  -0.45  

    \arrow <12pt> [.1,.4] from   0.5  0.0  to  0.5   2.000  
    \arrow <12pt> [.1,.4] from   1.0  0.0  to  1.0   1.000  
    \arrow <12pt> [.1,.4] from   1.5  0.0  to  1.5   0.667  
    \arrow <12pt> [.1,.4] from   2.0  0.0  to  2.0   0.250  
    \arrow <12pt> [.1,.4] from  -0.5  0.0  to  -0.5   -2.000  
    \arrow <12pt> [.1,.4] from  -1.0  0.0  to  -1.0   -1.000  
    \arrow <12pt> [.1,.4] from  -1.5  0.0  to  -1.5   -0.667  
    \arrow <12pt> [.1,.4] from  -2.0  0.0  to  -2.0   -0.250  



    \setquadratic    \setlinear

    \put {$\bullet$}  at 0.0 0.0  

    \put {$\bullet$}  at 0.5 0.0  
    \put {$\bullet$}  at 1.0 0.0  
    \put {$\bullet$}  at 1.5 0.0  
    \put {$\bullet$}  at 2.0 0.0  

    \put {{\large e}$_0$(0.5)} at 0.5 2.15
    \put {{\large e}$_0$(1.0)} at 1.0 1.15
    \put {{\large e}$_0$(1.5)} at 1.5 0.82
    \put {{\large e}$_0$(2.0)} at 2.0 0.40

\endpicture}
\vspace {100 pt}
$$
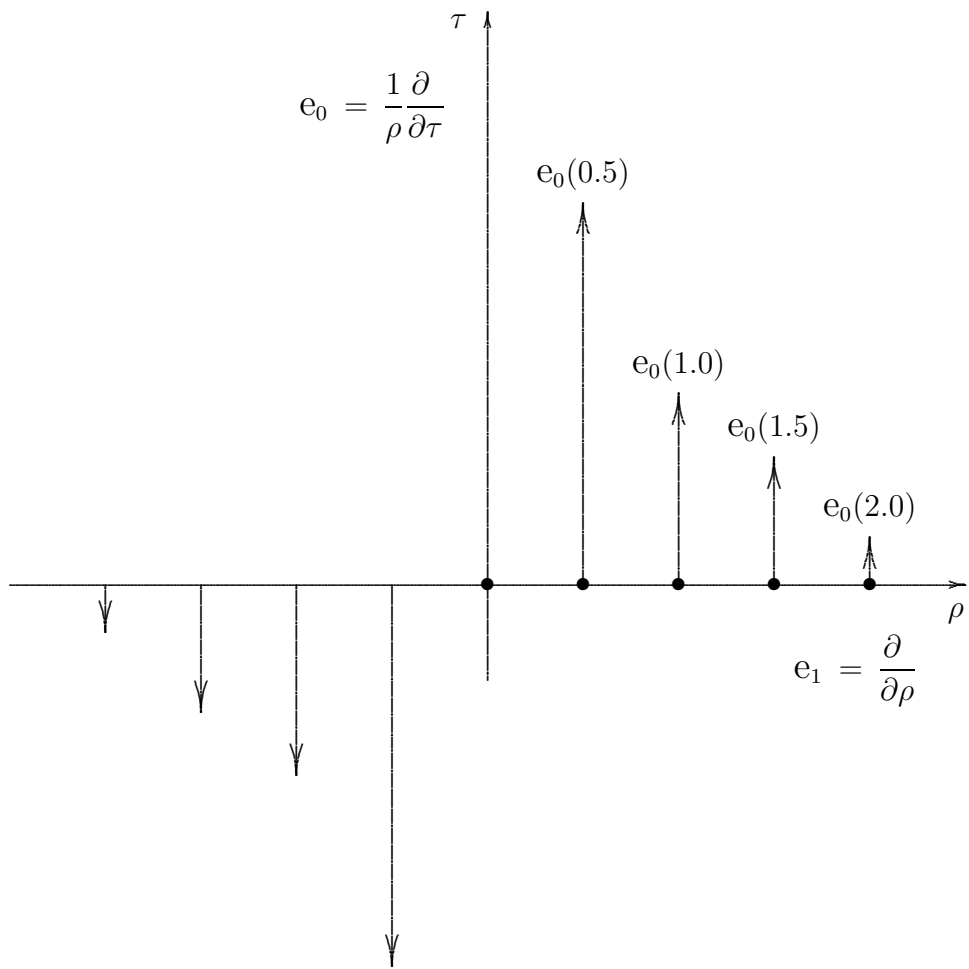
{\centerline {\bf Figure 0: Born Motion Vectors}} \newline


It is clear that the frame field is invariant under motions in the new time
coordinate $\tau$ but not under motions in the $\rho$-direction because
the size of the acceleration is $1/\rho$.

Here was a problem that Einstein apparently had not noticed in his 1907 paper
on the equivalence principle.
In Newton's theory the acceleration, $\gamma$, could be a vector
in the $x$-direction constant in space and time independent of
the velocity of a body moving in the $x$-direction.
Not so in Minkowski spacetime.
There the acceleration vector has to be orthogonal to the 4-velocity 
and it would appear that homogeneity of the acceleration field in spacetime
could no longer be achieved.

Apparently Max Planck had noticed the problem.
It was four months after Einstein had mailed his paper that
he sent a correction to the Jahrbuch.\cite{Einstein1908}
It began:
\begin{quote}
\lq\lq A letter by Mr.~Planck induced me to add the following
supplementary remark so as to prevent a misunderstanding that
could arise easily:
In the section \lq Principle of relativity and gravitation\rq,
a reference system at rest situated in a temporally constant, homogeneous
gravitational field is treated as  physically equivalent to a
uniformly accelerated, gravitation-free reference system.
The concept \lq uniformly accelerated\rq\ needs further clarification.\rq\rq
\end{quote}
Einstein then pointed out that the equivalence was to be restricted to a
body with zero velocity in the accelerated system.
In a linear approximation, he concluded, this was sufficient because
only linear terms had to be taken into account.

Einstein's retreat raises the question whether it is impossible
to find a homogeneous uniformly accelerated reference system, or,
assuming exact validity of his principle of equivalence,
a homogeneous gravitational field.

\section{A Homogeneous Gravitational Field}
\setcounter{equation}{0}

A definition of homogeneity in an $n$-dimensional manifold
involves the existence of $n$ linearly independent differential one-forms
\begin{equation} \label{eq:connforms}
{\mbox {\boldmath $\omega$}}'^{\mu}(x'^{\lambda})
\, = \,
\omega'^{\mu}{}_{\alpha}(x'^{\lambda}) \, {\bf dx}'^{\alpha}
\, , \qquad
\det \left[ \, \omega'^{\mu}{}_{\alpha} \, \right] \, \neq \, 0
\end{equation}
and $n$ independent functions
\begin{equation} \label{eq:coordf}
x'^{\lambda} \, = \, f^{\lambda}(x^{\mu}) \, , \qquad
\det \left[ \, \frac{\partial f^{\lambda}}{\partial x^{\mu}} \, \right] \, \neq \, 0 \, .
\end{equation}
The coordinate transformation Eq.(\ref{eq:coordf}) of the
differential forms Eq.(\ref{eq:connforms}), the so-called pull-back,
\begin{equation}
{\mbox {\boldmath $\omega$}}'^{\mu}(x'^{\lambda})
\, = \, \omega'^{\mu}{}_{\beta}(x'^{\lambda}) \, {\bf dx}'^{\beta}
\, = \, \omega'^{\mu}{}_{\beta}[ \, f^{\lambda}(x^{\nu}) \, ]
\frac{\partial f^{\beta}(x^{\nu})}{\partial x^{\alpha}} \, {\bf dx}^{\alpha}
\end{equation}
defines $n$ new differential forms $ {\mbox {\boldmath $\omega$}}^{\mu}(x^{\nu}) $ by
\begin{equation}
\omega'^{\mu}{}_{\beta}[ \, f^{\lambda}(x^{\nu}) \, ]
\frac{\partial f^{\beta}(x^{\nu})}{\partial x^{\alpha}} \, {\bf dx}^{\alpha}
\, \equiv \,
\omega^{\mu}{}_{\alpha}(x^{\nu}) \, {\bf dx}^{\alpha}
\, \equiv \,
{\mbox {\boldmath $\omega$}}^{\mu}(x^{\nu}) \, .
\end{equation}
If the $ \omega'^{\mu}{}_{\alpha} $ and $ \omega^{\mu}{}_{\alpha} $,
are the same functions of their arguments
\begin{equation} \label{eq:forminv}
\omega'^{\mu}{}_{\alpha}(x^{\nu}) \, = \, \omega^{\mu}{}_{\alpha}(x^{\nu}) \, ,
\end{equation}
we say the differential forms $ {\mbox {\boldmath $\omega$}}^{\mu} $ are invariant.

A simple example for Eq.(\ref{eq:forminv}) and Eq.(\ref{eq:coordf}) is
\begin{equation}
{\mbox {\boldmath $\omega$}}'^{\mu}(x'^{\lambda}) \, = \, {\bf dx}'^{\mu}
\, , \qquad
{\mbox {\boldmath $\omega$}}^{\mu}(x^{\nu}) \, = \, {\bf dx}^{\mu} \, , \qquad
x'^{\lambda} \, = \, x^{\lambda} \, + \, c^{\lambda} \, ,
\end{equation}
where $ c^{\lambda} \, = \, constant $, which immediately yields
\begin{equation}
\omega'^{\mu}{}_{\alpha}(x^{\nu}) \, = \,
\delta^{\mu}{}_{\alpha} \, = \,
\omega^{\mu}{}_{\alpha}(x^{\nu}) \, .
\end{equation}
The invariance, Eq.(\ref{eq:forminv}), of the differential forms gives 
\begin{equation}
\omega^{\mu}{}_{\alpha}(x'^{\lambda}) \,
\frac{\partial f^{\alpha}(x^{\nu})}{\partial x^{\beta}} \, {\bf dx}^{\beta}
\, = \,
\omega^{\mu}{}_{\beta}(x^{\nu}) \, {\bf dx}^{\beta} \, .
\end{equation}
Introducing the inverse of $ \omega^{\mu}{}_{\alpha} $ gives
\begin{equation} \label{eq:inverseforms1}
\xi^{\gamma}{}_{\mu}(x'^{\lambda}) \, \omega^{\mu}{}_{\alpha}(x'^{\lambda})
\, = \, \delta^{\gamma}{}_{\alpha}
\end{equation}
where
\begin{equation} \label{eq:inverseforms2}
\frac{\partial x'^{\gamma}}{\partial x^{\beta}}
\, = \,
\frac{\partial f^{\gamma}(x^{\nu})}{\partial x^{\beta}}
\, = \,
\xi^{\gamma}{}_{\mu}(x'^{\lambda}) \, \omega^{\mu}{}_{\beta}(x^{\nu}) \, .
\end{equation}
These are the equations of Lie's first fundamental theorem for Lie groups.

Differentiation of the $ {\mbox {\boldmath $\omega$}}^{\mu} $
gives the Maurer-Cartan equations
\begin{equation} \label{eq:maurer-cartan}
{\bf d}{\mbox {\boldmath $\omega$}}^{\mu}
\, + \,
\frac{1}{2} \, C^{\mu}{}_{\lambda\nu} \
{\mbox {\boldmath $\omega$}}^{\lambda} \wedge {\mbox {\boldmath $\omega$}}^{\nu}
\, = \, 0
\end{equation}
and the invariance of the $ {\mbox {\boldmath $\omega$}}^{\mu} $ leads to constancy of
the \lq\lq {\it structure constants}\rq\rq\ $ C^{\mu}{}_{\lambda\nu} $.
Further differentiation makes the left-hand side of Eq.(\ref{eq:maurer-cartan})
zero and creates thus the conditions for Lie's third fundamental theorem
of Lie groups by the integrability conditions
\begin{equation}
C^{\mu}{}_{\lambda\nu} \
{\bf d}{\mbox {\boldmath $\omega$}}^{\lambda} \wedge {\mbox {\boldmath $\omega$}}^{\nu}
\, - \,
C^{\mu}{}_{\lambda\nu} \
{\mbox {\boldmath $\omega$}}^{\lambda} \wedge {\bf d}{\mbox {\boldmath $\omega$}}^{\nu}
\, = \, 0 \, .
\end{equation}
With the Maurer-Cartan equations, this gives 
\begin{equation}
C^{\mu}{}_{\lambda\nu} \, C^{\nu}{}_{\rho\sigma} \
{\mbox {\boldmath $\omega$}}^{\nu}  \wedge
{\mbox {\boldmath $\omega$}}^{\rho} \wedge
{\mbox {\boldmath $\omega$}}^{\sigma}
\, = \, 0 \, .
\end{equation}
This gives the defining conditions for a Lie algebra
\begin{equation}
C^{\mu}{}_{\lambda\nu} \, C^{\nu}{}_{\rho\sigma}
\, + \,
C^{\mu}{}_{\rho\nu} \, C^{\nu}{}_{\sigma\lambda}
\, + \,
C^{\mu}{}_{\sigma\nu} \, C^{\nu}{}_{\lambda\rho}
\, = \, 0 \, .
\end{equation}
By introducing Lie groups through invariant differential forms,
\'Elie Cartan was able to simplify Lie's derivations of his fundamental theorems.

A Riemannian manifold is called {\it homogeneous} if its metric is
invariant under a transitive group of motions.
Mathematicians call a transformation group {\it transitive}
if its action maps any point of the manifold into any other point.
The $ \mbox{\bf S}^{2} $, for example,
with its standard metric inherited from being
rigidly embedded into the 3-dimensional Euclidean space,
is homogeneous under the action of the rotation group $ \mbox{\bf O}(3) $.

The homogeneous manifolds we need for gravitational fields are of a special nature.
For them the group of motions has to be {\it simply transitive},
meaning that there is only one group element that moves a
point of the manifold into another given point of the manifold.
In this case a Riemannian manifold is homogeneous
if the metric 
\begin{equation}
ds^{2} \, = \,
\eta_{\mu\nu} \, {\mbox {\boldmath $\omega$}}^{\mu} \, {\mbox {\boldmath $\omega$}}^{\nu} \, , \qquad
\det \left[ \, \eta_{\mu\nu} \, \right] \, \neq \, 0 \, , \qquad
\eta_{\mu\nu} \, = \, \eta_{\nu\mu} \, = \, constant \, ,
\end{equation}
is given in terms of invariant differential one-forms
$ {\mbox {\boldmath $\omega$}}^{\mu}(x^{\lambda}) $
for a simply transitive group.
These one-forms define in each point of the manifold an orthonormal $n$-leg
of vectors, also known as a \lq\lq frame\rq\rq\ by
\begin{equation} \label{eq:HGFomega(e)}
{\mbox {\boldmath $\omega$}}^{\mu}({\bf e}_{\lambda}) \, = \, \delta^{\mu}{}_{\lambda} \, .
\end{equation}
In coordinate components, these vectors are then given by
Eq.(\ref{eq:inverseforms1}) as
\begin{equation}
{\bf e}_{\lambda} \, \longrightarrow \, \xi^{\alpha}{}_{\lambda} \, .
\end{equation}

The Levi-Civita connection is defined by differential one-forms
$ {\mbox {\boldmath $\omega$}}^{\mu}{}_{\nu} $,
the {\it connection forms},
that represent the gravitational forces given by
\begin{equation}
{\bf d}{\mbox {\boldmath $\omega$}}^{\mu} \, = \,
{\mbox {\boldmath $\omega$}}^{\mu}{}_{\nu} \wedge {\mbox {\boldmath $\omega$}}^{\nu}
\, , \qquad
{\mbox {\boldmath $\omega$}}_{\mu\nu} \, = \, - \, {\mbox {\boldmath $\omega$}}_{\nu\mu} \, .
\end{equation}
These are \'Elie Cartan's first structural equations for a Riemannian
space with vanishing torsion.
Some authors define the connection forms
$ {\mbox {\boldmath $\omega$}}^{\mu}{}_{\nu} $ with
the opposite sign from our convention.

The components of the gravitational field, $ g^{\mu}{}_{\nu\lambda} $, are given by
\begin{equation} \label{eq:gfcomponents}
{\mbox {\boldmath $\omega$}}^{\mu}{}_{\nu} \, = \,
g^{\mu}{}_{\nu\lambda} \, {\mbox {\boldmath $\omega$}}^{\lambda} \, , \qquad
g_{\mu\nu\lambda} \, + \, g_{\nu\mu\lambda} \, = \, 0 \, .
\end{equation}
We have thus
\begin{equation}
{\bf d}{\mbox {\boldmath $\omega$}}^{\mu} \, = \,
g^{\mu}{}_{\nu\lambda} \
{\mbox {\boldmath $\omega$}}^{\lambda} \wedge {\mbox {\boldmath $\omega$}}^{\nu} \, .
\end{equation}
Comparing this equation with the Maurer-Cartan equation with the indices
lowered by $ \eta_{\mu\rho} $ gives
\begin{equation}
{\bf d}{\mbox {\boldmath $\omega$}}_{\alpha} \, = \, \frac{1}{2}
\left( g_{\alpha\lambda\nu} \, -  \, g_{\alpha\nu\lambda} \right) \
{\mbox {\boldmath $\omega$}}^{\nu} \wedge {\mbox {\boldmath $\omega$}}^{\lambda}
\, = \, - \, \frac{1}{2} \, C_{\alpha\nu\lambda} \
{\mbox {\boldmath $\omega$}}^{\nu} \wedge {\mbox {\boldmath $\omega$}}^{\lambda}
\, .
\end{equation}
We obtain, therefore, with even permutations, 
\begin{equation} \label{eq:cyclic1}
C_{\alpha\lambda\nu} \, = \, g_{\alpha\lambda\nu} \, -  \, g_{\alpha\nu\lambda} \, ,
\end{equation}
\begin{equation} \label{eq:cyclic2}
C_{\lambda\nu\alpha} \, = \, g_{\lambda\nu\alpha} \, -  \, g_{\lambda\alpha\nu} \, ,
\end{equation}
\begin{equation} \label{eq:cyclic3}
C_{\nu\alpha\lambda} \, = \, g_{\nu\alpha\lambda} \, -  \, g_{\nu\lambda\alpha} \, .
\end{equation}
Adding Eq.(\ref{eq:cyclic1}) and Eq.(\ref{eq:cyclic2}) and
subtracting Eq.(\ref{eq:cyclic3}) gives, because of the skew-symmetry
of $ g_{\mu\nu\lambda} $ in its first two indices according to Eq.(\ref{eq:gfcomponents}),
that
\begin{equation} \label{eq:gccc}
2 \, g_{\alpha\lambda\nu}
\, = \,
C_{\alpha\lambda\nu} \, + \, C_{\lambda\nu\alpha} \, - \, C_{\nu\alpha\lambda} \, .
\end{equation}
This shows that the physical components of a gravitational field's strength
are constant in a homogeneous gravitational field.

For the further development it is useful to discuss an example of such
a homogeneous gravitational field.

\section{A Homogeneous Gravitational Field in the Minkowski Plane}
\setcounter{equation}{0}

It is easy to find invariant differential forms if one writes the metric
for the Minkowski plane as
\begin{equation} \label{eq:HGFMinkPlaneEta}
ds^{2}
\, = \, \eta_{\mu\nu} \, dx^{\mu} \, dx^{\nu}
\, = \, 2 \, du \, dv \, , \qquad
\eta_{\mu\nu} \, = \, \left( \begin{array}{lr} 0 \ 1 \\ 1 \ 0 \end{array} \right) .
\end{equation}
where we have used the null coordinates
\begin{equation} \label{eq:utxvtx}
u \, = \, \frac{1}{\sqrt 2} \, ( \, t \, - \, x \, ) \, , \qquad
v \, = \, \frac{1}{\sqrt 2} \, ( \, t \, + \, x \, ) \, ,
\end{equation}
and the associated vectors
\begin{equation}
\frac{\partial \ }{\partial u} \, = \, \frac{1}{\sqrt 2}
\left( \frac{\partial \ }{\partial t} \, - \,
\frac{\partial \ }{\partial x} \right) \, , \qquad
\frac{\partial \ }{\partial v} \, = \, \frac{1}{\sqrt 2}
\left( \frac{\partial \ }{\partial t} \, + \,
\frac{\partial \ }{\partial x} \right) \, .
\end{equation}
The one-forms
\begin{equation}
{\mbox {\boldmath $\omega$}}^{0} \, = \, \frac{du}{\gamma \, u} \, , \qquad
{\mbox {\boldmath $\omega$}}^{1} \, = \, \gamma \, u \, dv \, , \qquad
ds^{2}
\, = \,
2 \, {\mbox {\boldmath $\omega$}}^{0} \, {\mbox {\boldmath $\omega$}}^{1} \, , \qquad
\gamma \, = \, constant \, \neq \, 0 \, ,
\end{equation}
are independent and give us constant structure coefficients.
We have
\begin{equation}
d{\mbox {\boldmath $\omega$}}^{0}
\, = \, 0 \, = \, 
- \, C^{0}{}_{01} \,
{\mbox {\boldmath $\omega$}}^{0} \wedge {\mbox {\boldmath $\omega$}}^{1} \, ,
\end{equation}
\begin{equation}
d{\mbox {\boldmath $\omega$}}^{1}
\, = \, \gamma \ du \wedge dv
\, = \, \gamma \ {\mbox {\boldmath $\omega$}}^{0} \wedge {\mbox {\boldmath $\omega$}}^{1}
\, = \, - \, C^{1}{}_{01} \ 
{\mbox {\boldmath $\omega$}}^{0} \wedge {\mbox {\boldmath $\omega$}}^{1}
\end{equation}
which gives
\begin{equation} \label{eq:MC001C101}
C^{0}{}_{01} \, = \, 0 \, , \qquad C^{1}{}_{01} \, = \, - \, \gamma \, .
\end{equation}
We then have, from Eq.(\ref{eq:gccc}), for the gravitational field strength(s)
\begin{equation}
g_{100} \, = \, - \, \gamma \, , \qquad g_{101} \, = \, 0 \, .
\end{equation}
The frame vectors are given by Eq.(\ref{eq:HGFomega(e)});
in the current situation,
they are 
\begin{equation} \label{eq:Me0e1}
{\bf e}_{0} \, = \, \gamma \, u \, \frac{\partial \ }{\partial u} \, , \qquad
{\bf e}_{1} \, = \, \frac{1}{\gamma \, u} \, \frac{\partial \ }{\partial v} \, .
\end{equation}
A general velocity vector in the Minkowski plane is subject to
\begin{equation} \label{eq:v0v1}
2 \, V^{0} \, V^{1} \, = \, 1
\end{equation}
and is thus time-like.
It is directed into the future if $ V^{0} > 0 $.

The acceleration vector $ {\bf g} $ has the covariant components $ g_{j} $ equal to
\begin{equation} \label{eq:g_j}
g_{j} \, = \, g_{jkl} \, V^{k} \, V^{l} \, , 
\end{equation}
which is the geodesic acceleration adapted to our frame.
In our case, we obtain
\begin{equation} \label{eq:g0g1}
g_{0} \, = \, g_{010} \, V^{1} \, V^{0} \, = \, \gamma \, V^{1} \, V^{0} \, , \qquad
g_{1} \, = \, g_{100} \, V^{0} \, V^{0} \, = \, - \, \gamma \, \left( V^{0} \right)^{2} \, . 
\end{equation}
Evidently, this expression for the acceleration fulfills the requirement that
it be orthogonal to the 4-velocity; that is,
\begin{equation}
g_{j} \, V^{j} \, = \, 0 \, .
\end{equation}
If we choose the velocity as a constant, that is, $ V^{0} \, = \, constant \, > \, 0 $,
it follows from Eq.(\ref{eq:v0v1}) that $ V^{1} $ is also constant and the velocity
vector field
\begin{equation} \label{eq:MinkVvector}
{\bf V} \, = \, V^{\mu} \, {\bf e}_{\mu}
\, = \,
V^{0} \, \gamma \, u \, \frac{\partial \ }{\partial u} \, +
\frac{1}{2 \, V^{0} \, \gamma \, u} \, \frac{\partial \ }{\partial v} \, .
\end{equation}
is invariant in the Minkowski plane.
To better visualize the field, we calculate the worldlines that
have the vectors of the field as tangents.
We have
\begin{equation}
\frac{du}{ds} \, = \, V^{0} \, \gamma \, u \, , \qquad
\frac{dv}{ds} \, = \, \frac{1}{2 \, V^{0} \, \gamma \, u} \,
\end{equation}
or, with integration constants $ u_{0}, v_{0} $, we get
\begin{equation}
u \, = \, u_{0} \, e^{V^{0} \, \gamma \, s} \, , \qquad
v \, - \, v_{0} \, = \,
- \, \frac{e^{- \, V^{0} \, \gamma \, s}}{2 \, u_{0} \, (V^{0} \, \gamma)^{2}} \, .
\end{equation}
By eliminating $ u_{0} $ and $ s $,
we obtain the one-parameter set of curves 
\begin{equation} \label{eq:v-v0}
v \, - \, v_{0} \, = \, - \, \frac{1}{2 \, u \, (V^{0} \, \gamma)^{2}} \, .
\end{equation}
This is a set of identical hyperbolae that are obtained from the hyperbola
\begin{equation} \label{eq:uv}
u \, v \, + \, \frac{1}{2 \, (V^{0} \, \gamma)^{2}} \, = \, 0
\end{equation}
by translation in the $v$-direction.
With Eq.(\ref{eq:utxvtx}), this hyperbola can also be written as
\begin{equation} \label{eq:t2x2}
t^{2} \, - \, x ^{2} \, + \, \frac{1}{(V^{0} \, \gamma)^{2}} \, = \, 0  \, .
\end{equation}
[{\bf See Figure~1 ; left and right hyperbolae.}]

$$
\hskip 0.0 in
{\beginpicture
    \setcoordinatesystem units <.25in,.25in> point at 0 0

    \arrow <12pt> [.1,.4] from -10.0    0.0   to  10.0   0.0  
    \arrow <12pt> [.1,.4] from   0.0  -10.0   to   0.0  11.0  

    \arrow <12pt> [.1,.4] from   -10.0   -10.0  to  10.0   10.0  
    \arrow <12pt> [.1,.4] from    10.0   -10.0  to -10.0   10.0  

    \put {$ \odot $}  at  5.5  0.0
    \put {$ \odot $}  at -5.5  0.0

    \setquadratic

    \plot  10  -9.5    5.5  0    10  9.5  /  
    \plot -10  -9.5   -5.5  0   -10  9.5  /  

    \setlinear

    \put {$ \displaystyle  x = \frac{1}{V^{0} \gamma} $}      at  4.2  -0.8

    \put {$ \displaystyle  x = - \, \frac{1}{V^{0} \gamma} $} at -7.2 -0.8


    \put {$ \odot $}  at  0.0  5.5
    \put {$ \odot $}  at  0.0 -5.5

    \setquadratic

    \plot -10  10.5   0  5.5   10  10.5  /  
    \plot -10 -10.5   0 -5.5   10 -10.5  /  

    \setlinear

    \put {$ \displaystyle  y = \frac{1}{V^{0} \gamma} $} at 1.3 4.7

    \put {$ \displaystyle  y = - \, \frac{1}{V^{0} \gamma} $} at -1.5 -6.6


    \put {\large u} at -6.0 6.5
    \put {\large v} at  6.0 6.5

    \put {$\bullet$} at 0 0  

    \put {\large t} at -0.3 10.0
    \put {\large x} at  9.0  0.3

\endpicture}
\vspace {80 pt}
$$
{\centerline {\bf Figure 1: A Homogeneous Gravitational Field in the Minkowski Plane}} \newline

Next, we calculate the acceleration vector field.
We have from Eq.(\ref{eq:g0g1})
\begin{equation}
g^{0} \, = \, - \, \gamma \, (V^{0})^{2} \, , \qquad
g^{1} \, = \, \gamma \, V^{1} \, V^{0}
\end{equation}
and thus
\begin{equation}
{\bf g} \, = \, g^{\mu} \, {\bf e}_{\mu}
\, = \, - \, \gamma^{2} \, (V^{0})^{2} \, u \, \frac{\partial \ }{\partial u} \, +
\frac{V^{1} \, V^{0}}{u} \, \frac{\partial \ }{\partial v} \, .
\end{equation}
This vector field is tangent to the curves
\begin{equation}
\frac{du}{d\tau} \, = \, - \, (\gamma \, V^{0})^{2} \, u \, , \qquad
\frac{dv}{d\tau} \, = \, \frac{V^{1} \, V^{0}}{u} 
\end{equation}
described by the parameter $\tau$.
We obtain
\begin{equation}
u \, = \, u_{1} \, e^{- ( \gamma V^{0} )^{2} \tau} \, , \qquad
v \, - \, v_{1}
\, = \,
\frac{V^{1} \, V^{0}}{( \, \gamma \, V^{0} \,)^{2} \, u_{1}} \, e^{( \gamma V^{0} )^{2} \tau}
\end{equation}
with integration constants $ u_{1}, v_{1} $.
By eliminating $ u_{1} $ and $ v_{1} $ together with Eq.(\ref{eq:v0v1}),
we obtain
\begin{equation}
v \, - \, v_{1} \, = \, \frac{1}{2 \, u \, ( \, \gamma \, V^{0} \, )^{2}} \, .
\end{equation}
These curves are obtained from the hyperbola
\begin{equation}
u \, v \, = \, \frac{1}{2 \, ( \, \gamma \, V^{0} \,)^{2}}
\end{equation}
by translation in the $v$-direction.
With Eq.(\ref{eq:utxvtx}), this hyperbola can also be written
\begin{equation}
t^{2} \, - \, x ^{2} \, = \, \frac{1}{( \, \gamma \, V^{0} \,)^{2}} \, .
\end{equation}
[{\bf See Figure~1 ; upper and lower hyperbolae.}]
By shifting the lower branch of the hyperbola by $ \sqrt{2}/V^{0}\gamma $
in the positive $v$-direction, we intersect then the time-like hyperbola
Eq.(\ref{eq:uv}) at $ t = 0 $, $ x = 1/V^{0}\gamma $.
[{\bf See Figure~2.}]
Tangent to the hyperbolae at their intersection are the two frame vectors
of the velocity field and its acceleration.
Now, we wish to show that in the neighborhood of the intersection point
we are approximating a Newtonian homogeneous gravitational field.

\section{The Newtonian Gravitational Field}
\setcounter{equation}{0}

We study the Newtonian field with acceleration $-\gamma$ in the $x$-direction
of zero velocity and at $ t = 0 $.
In the Minkowski $t$-$x$--plane, we take parallel worldlines $ x = constant $
that are geodesics.
For $ t = 0 $ and $ x = 1 / V^{0}\gamma $, the parallel to the $t$-axis through
this point is touched by the hyperbola
Eq.(\ref{eq:t2x2}).
This hyperbola, parameterized by with proper time $s$,
\begin{equation}
t \, = \, \rho \ \sinh \frac{s}{\rho} \, , \qquad
x \, = \, \rho \ \cosh \frac{s}{\rho} \, , \qquad
\rho \, =\, \frac{1}{V^{0} \, \gamma} \,
\end{equation}
is the worldline of an observer with constant intrinsic acceleration.
Now, let this observer measure the distance to the straight line $ x = \rho $
orthogonal to his worldline.
This distance is a function $ \eta(s) $.
From
{\bf Figure~3},
we read off immediately that 
\begin{equation}
\eta \, = \, \rho \, - \, \frac{\rho}{\cosh(s/\rho)}
\, = \,
\rho \ \frac{\cosh(s/\rho)-1}{\cosh(s/\rho)}
\, = \, \frac{1}{2 \rho} \, s^{2} \, + \, \dots
\end{equation}
where higher terms are of fourth-order in time $s$.
The acceleration is given by
\begin{equation}
a \, = \, \frac{d^{2}\eta}{ds^{2}} \, = \, \frac{1}{\rho} \, = \, V^{0} \, \gamma \, .
\end{equation}
Since the velocity $ d\eta/ds $ increases,
the acceleration points into the negative $x$-direction.

$$
\hskip 0.5in
\beginpicture
    \setcoordinatesystem units <.25in,.25in> point at 0 0

    \arrow <12pt> [.1,.4] from  -5.0    0.0   to  12.0   0.0  
    \arrow <12pt> [.1,.4] from   0.0   -9.0   to   0.0  11.0  

    \arrow <12pt> [.1,.4] from  -5.0   -5.0  to  10.0   10.0  
    \arrow <12pt> [.1,.4] from   9.0   -9.0  to  -5.0    5.0  

    \put {$ \odot $}  at  5.5  0.0


    \setquadratic

     \plot  -5.00  -5.50   -3.00  -3.60    -1.00  -2.11   /  
     \plot  -1.00  -2.11    5.50   0.00    12.00  -2.11   /  
     \plot  -5.00  -5.53   -3.00  -3.65    -1.00  -2.13   /  
     \plot  -1.00  -2.13    5.50  -0.03    12.00  -2.13   /  

    \plot  9.555   -9.0    5.5  0.0    9.555   9.0  /  
    \plot  9.585   -9.0    5.53  0.0   9.585   9.0  /  

    \plot  9.6  3.8    6.9 1.6    5.5 0.0  /  

    \setlinear


    \put {$ \displaystyle  x = \frac{1}{V^{0} \gamma} $} at  10.0  4.0

    \put {\large u} at -3.5 4.0
    \put {\large v} at  8.5 9.0
    \put {$\bullet$}  at 0 0  
    \put {\large t} at -0.3 10.0
    \put {\large x} at  11.0  0.3

\endpicture
$$
\vskip 2.8in
{\centerline {
\bf Figure 2: A Homogeneous Gravitational Field in the Minkowski Plane}}
\medskip
{
\noindent
\hskip 0.0in
\vbox{\noindent
The two hyperbolic branches intersecting at $ x = 1 / ( V^0\gamma ) $
are the right branch of the hyperbola in Figure 1 and lower branch of
the hyperbola in Figure 1 displaced in the $v$-direction.
The two curves intersect orthogonally and have as their unit tangent vectors
$ {\bf e}_{0} $ and $ {\bf e}_{1} $ of a homogeneous gravitational field.
}}
\medskip
\noindent 
\hrulefill
\medskip


It is for this special case: 
against absolute space, in the same event, and at the same velocity,
an inertial observer and the accelerated one
can interpret gravity as acceleration and vice versa.
Here the Newtonian equivalence for a homogeneous gravitational field
can be made relativistic in a point.
This raises the question: Is it possible to save Einstein's
equivalence for a homogeneous gravitational field.

$$
\hskip 0.5 in
{\beginpicture
    \setcoordinatesystem units <.25in,.25in> point at 0 0

    \arrow <12pt> [.1,.4] from  -5.0    0.0   to  12.0   0.0  
    \arrow <12pt> [.1,.4] from   0.0  -10.0   to   0.0  11.0  

    \arrow <12pt> [.1,.4] from    -5.0    -5.0  to  10.0   10.0  
    \arrow <12pt> [.1,.4] from    10.0   -10.0  to  -5.0    5.0  

    \put {$ \odot $}  at  5.5  0.0

    \put {$ \odot $}  at  5.5  4.64
    \arrow <12pt> [.1,.4] from   5.5  4.64  to  7.5 6.34   
    \plot 5.5  4.67   7.5 6.36     /  
    \plot 5.5  4.61   7.5 6.32     /  

    \setquadratic

    \plot  10.0   -9.5    5.5   0.0    10.0    9.5  /  
    \plot  10.03  -9.5    5.53  0.0    10.03   9.5  /  

    \plot   5.5 0.0   5.95 3.1   7.5 6.34   /  %

    \plot   0.0   0.0   0.75 -0.4   1.5  -0.3   2.30 -0.2   2.75 -0.5  /  
    \plot   2.75 -0.5   3.20 -0.2   4.0  -0.3   4.75 -0.4   5.5   0.0  /  

    \plot   0.0 0.0     0.55  1.5   2.0  2.6     2.8 3.4     2.9 4.6    /  
    \plot   2.9 4.6     3.8   4.5   4.9  5.45    6.2 6.5     7.5 6.34   /  

    \plot   7.50  0.00  7.9 0.500    7.8 1.40  7.7 2.285  8.25  3.17   /  %
    \plot   8.25  3.17  7.7 4.055    7.8 4.94  7.9 5.840  7.50  6.34   /  %

    \setlinear 

    \setdashes
    \plot 5.5  -10.0   5.5 11.0   /  %
    \setsolid

    \plot 0.0  0.0   7.5 6.34   /  %
    \plot 7.5  0.0   7.5 6.34   /  

    \put {\large u} at -3.5 4.0
    \put {\large v} at  8.5 9.0
    \put {$\bullet$}  at 0 0  
    \put {\large t} at -0.3 10.0
    \put {\large x} at  11.0  0.3

    \put {s} at  6.250  2.8

    \put {$\eta$} at 6.1 5.5

    \put {$\rho$} at 2.75 -0.8
    \put {$\rho$} at 2.75  4.8

    \put {$ \rho \cosh s/\rho $} at 7.50 -0.8
    \put {$ \rho \sinh s/\rho $} at 9.50  3.17

    \put {$\rho = \frac{1}{V^{0}\gamma} $} at 2.75 -5.8

\endpicture}
$$
\vskip 1.50in
{\centerline {\bf Figure 3: The Newtonian Gravitational Field}}
\medskip
\noindent
The worldline described by the parametric equations
$$
x \, = \, \rho \, \cosh \frac{s}{\rho} \, , \qquad
t \, = \, \rho \, \sinh \frac{s}{\rho} \, , \qquad
\rho \, = \, \frac{1}{V^{0}\gamma}
$$
has acceleration $ 1 / \rho $.
It osculates, at $ t = 0 $, the geodesic $ \rho = constant $.
The vector $ \eta $ determines the distance from the geodesic
to the hyperbola in the rest system of the accelerated observer,
that is, \lq\lq orthogonal\rq\rq\ to his worldline.
\vfill

\section{Relativistic Equivalence}
\setcounter{equation}{0}

We have applied the calculus of Ricci Curbastro to Minkowski space-time
for the description of a homogeneous gravitational field in Einstein's theory.
It turns out that such fields actually exist, without any space-time curvature,
in a flat world.
The existence of such a non-vanishing field came as a surprise to us because
its analogs on the $ {\bf S}^{2} $ or the Euclidean plane do not exist.
The reason for its existence on a flat space-time is that the Poincar\'e group
in (1,1)-dimensions has a simply transitive 2-dimensional subgroup.
This is not the case for {\bf SO(3)}.
The Euclidean motions in the plane, {\bf E(2)},
only have the trivial transformations.

In terms of the light-like coordinates $u$ and $v$,
the Poincar\'e group of the Minkowski plane can be written
\begin{equation}
u' \, = \, \alpha \, u \, + \, a \, , \qquad
v' \, = \, \frac{1}{\alpha} \, v \, + \, b \, , \qquad
\alpha \, \neq \, 0
\end{equation}
with constants $a$, $b$, $\alpha$.
In matrix form, this is
\begin{equation}
\left(
\begin{array}{c}
u' \\
v' \\
1
\end{array}
\right)
=
\left(
\begin{array}{ccc}
\alpha & 0 & a \\
0 & \alpha^{-1} & b \\
0 & 0 & 1
\end{array}
\right)
\left(
\begin{array}{c}
u \\
v \\
1
\end{array}
\right) \, .
\end{equation}
Clearly, the matrices with vanishing $a$
\begin{equation} \label{eq:subgroup}
\left(
\begin{array}{ccc}
\alpha & 0 & 0 \\
0 & \alpha^{-1} & b \\
0 & 0 & 1
\end{array}
\right)
\end{equation}
form a subgroup of the Poincar\'e group depending on
the two parameters $\alpha$ and $b$.
For events with $ u' \neq 0 $, a unique event with
coordinates $ ( u, v ) $ exists and is obtained from the inverse matrix of
Eq.(\ref{eq:subgroup}).
This demonstrates that the subgroup acts simply transitively in the two
demi-mondes $ u \neq 0 $.
The existence of this subgroup is the source of the homogeneous field.
But it is also the source of what is known as \lq\lq teleparallelism\rq\rq.

If we take an orthonormal frame in one original point with say $ u < 0 $,
the motions of the subgroup move it to all other points with $ u < 0 $
in a unique fashion.
When we assign to a vector $ {\bf V} $ in the original point
the components $ V^{0} $ and $ V^{1} $,
we can now assign these same components to vectors with
respect to the frames that were transported and now define them
as parallel to each other.
This is clearly what mathematicians call an equivalence relation
known as teleparallelism.

Such a more general notion of parallelity,
based on simply transitive groups,
was introduced by the Princeton geometer and later Dean
Luther Pfahler Eisenhart in a brief paper in 1925.\cite{Eisenhart}
The concept was further developed by \'Elie Cartan
together with the Dutch electrical engineer Jan Arnoldus Schouten
who had discovered parallel displacement in Riemannian geometry
independently of Levi-Civita.\cite{CartanSchouten1926}

The homogeneity group is characterized by the invariant differential forms
$ {\mbox {\boldmath $\omega$}}^{\mu} $ of Eq.(\ref{eq:connforms})
that determine the frames.
We get the frames everywhere if we pick a frame in one point and
integrate the total differential equations
Eq.(\ref{eq:maurer-cartan})
of Ludwig Maurer and \'Elie Cartan.
We call the vectorial 2-form
\begin{equation} \label{eq:ThetaDef}
{\bf \Theta}^{\mu} \, = \, d{\mbox {\boldmath $\omega$}}^{\mu} \, = \,
- \, \frac{1}{2} \, C^{\mu}{}_{\lambda\nu} \ 
{\mbox {\boldmath $\omega$}}^{\lambda} \wedge {\mbox {\boldmath $\omega$}}^{\nu}
\end{equation}
the torsion form.
In any case, it is simply determined by the structure constants, $ C^{\mu}{}_{\lambda\nu} $,
of the homogeneity group.

\section{Cartan's Torsion}
\setcounter{equation}{0}

\'Elie Cartan introduced the notion of torsion in a brief note in 1922.\cite{Cartan1922}
He amplified his sketch in a memoir in the following year.
It was reprinted in his collected papers in 1955.
This paper {\lq\lq {\it Sur les variet\'es \`a connexion affine et la th\'eorie
de la relativit\'e g\'en\'eralis\'ee\rq\rq}} \,
is now accessible through its translation in book form and an introduction
that facilitates its study.\cite{Cartan1923, Cartan1924}

It is easiest to introduce torsion by using the notion of the exterior
covariant derivative of a vector-valued differential one-form.
For the general formalism see, for example, EDM2.\cite{EDM2}

In the tangent vector space attached to each point of a manifold
exists a unit operator $I$.
A representation of it in terms of $n$ real Dirac kets $|\,i>$ and
bras $<j\,|$ would be given by
\begin{equation}
I \, = \, |\,\mu><\mu\,| \, , \qquad \mu \, \in \, 1, \, \dots , \, n
\end{equation}
with
\begin{equation}
< \nu \, | \, I \, | \, \mu > \ \ = \ \ < \nu \, | \, \mu > \ \ = \ \ \delta_{\nu\mu}
\end{equation}
indicating orthonormality.
The torsion is described by the vector-valued 2-form ${\bf \Theta}$ as
\begin{equation}
{\bf \Theta} \, = \, \nabla I \, .
\end{equation}
Here, $\nabla$ indicates the operator of exterior covariant derivation.

Since the Dirac notation is not used in differential geometry,
one writes rather
\begin{equation}
I \, = \, {\bf e}_{\mu} \otimes {\mbox {\boldmath $\omega$}}^{\mu} \, , \qquad
{\mbox {\boldmath $\omega$}}^{\nu} \left( {\bf e}_{\mu} \right)
\, = \, \delta^{\nu}{}_{\mu} \, .
\end{equation}
The connection is then defined by
\begin{equation} \label{eq:codivvect}
\nabla {\bf e}_{\mu} \, = \, - \, {\bf e}_{\nu} \, {\mbox {\boldmath $\omega$}}^{\nu}{}_{\mu}
\end{equation}
where the $ {\mbox {\boldmath $\omega$}}^{\nu}{}_{\mu} $ are the connection 1-forms.

The torsion ${\bf \Theta}$ becomes then 
\begin{equation}
{\bf \Theta}
\, = \,
\nabla \left( {\bf e}_{\mu} \, {\mbox {\boldmath $\omega$}}^{\mu} \right)
\, = \,
- \,{\bf e}_{\nu} \,
{\mbox {\boldmath $\omega$}}^{\nu}{}_{\mu} \wedge {\mbox {\boldmath $\omega$}}^{\mu}
\, + \,
{\bf e}_{\nu} \, d{\mbox {\boldmath $\omega$}}^{\nu}
\, = \,
{\bf e}_{\nu} \left( \, d{\mbox {\boldmath $\omega$}}^{\nu}
\, - \,
{\mbox {\boldmath $\omega$}}^{\nu}{}_{\mu} \wedge {\mbox {\boldmath $\omega$}}^{\mu} \, \right) \, .
\end{equation}
Here, we are interested only in affinely connected manifolds
for which Eq.(\ref{eq:codivvect}) holds and where the connection
leaves the metric invariant, that is, the scalar product of vectors
in the tangent space.
We have
\begin{eqnarray}
\nabla \left( \, {\bf e}_{\lambda} \cdot {\bf e}_{\mu} \, \right)
&\, = \,& \nabla {\bf e}_{\lambda} \cdot {\bf e}_{\mu}
\, + \, {\bf e}_{\lambda} \cdot \nabla {\bf e}_{\mu} \nonumber \\
&\, = \,&  - \, {\bf e}_{\nu} \cdot {\bf e}_{\mu} \
{\mbox {\boldmath $\omega$}}^{\nu}{}_{\lambda}
\, - \, {\bf e}_{\lambda} \cdot {\bf e}_{\nu} \
{\mbox {\boldmath $\omega$}}^{\nu}{}_{\mu} \nonumber \\
&\, = \,& - \, \left( \, {\mbox {\boldmath $\omega$}}_{\mu\lambda}
\, + \,
{\mbox {\boldmath $\omega$}}_{\lambda\mu} \, \right) \nonumber \\
&\, = \,& \, \nabla \left( \eta_{\lambda\mu} \right) \nonumber \\
&\, = \,& 0 \, .
\end{eqnarray}
This means that the connection form with covariant indices
must be skew in these indices.

We now want to study the constant torsion in the Minkowski plane.
Writing the vector components of the torsion 2-form
\begin{equation}
{\bf \Theta}^{\mu} \, = \,
\frac{1}{2} \, T^{\mu}{}_{\lambda\nu}
\ {\mbox {\boldmath $\omega$}}^{\lambda} \wedge {\mbox {\boldmath $\omega$}}^{\nu} \, ,
\end{equation}
we have, from Eq.(\ref{eq:ThetaDef}) and Eq.(\ref{eq:MC001C101})
\begin{equation}
T^{0}{}_{01} \, = \, 0 \, , \qquad T^{1}{}_{01} \, = \, \gamma \, .
\end{equation}
This shows that the torsion is constant and thus homogeneous.

We now want to look at the basis vectors
$ {\bf e}_{0} $ and $ {\bf e}_{1} $
of our distant parallelism.
We have from Eq.(\ref{eq:Me0e1})
\begin{equation} \label{eq:Me0e1X}
{\bf e}_{0} \, = \, \gamma \, u \, \partial_{\mu} \, , \qquad
{\bf e}_{1} \, = \, \frac{1}{\gamma \, u} \, \partial_{v} \, .
\end{equation}
We take a point $P$ on the line $ \gamma \, u \, = -1 $
and draw the frame vectors $ {\bf e}_{0}(P) $ and $ {\bf e}_{1}(P) $.
The vector $ {\bf e}_{0}(P) $ has its tip in the point $Q$
that lies on the line $ \gamma \, u \ = -2 $.
According to Eq.(\ref{eq:Me0e1X}), the vector $ {\bf e}_{1}(Q) $ is  only
half as long as the vector $ {\bf e}_{1}(P) $.
[{\bf See Figure~4.}]

We now assume that the connection coefficients vanish:

The vector $ {\bf e}_{0}(P') $,
issuing from the tip of the vector $ {\bf e}_{1}(P) $,
is obtained through parallel transfer from $ {\bf e}_{0}(P) $ and
has remained unchanged since $ \gamma \, u $ was constant.
The effect of the torsion is that our square $PQP'Q'$ does not close.
There should be a physical explanation for this phenomenon.

$$
\hskip 0.0 in
{\beginpicture
    \setcoordinatesystem units <.25in,.25in> point at 0 0

    \arrow <12pt> [.1,.4] from   -5.0   -5.0  to  10.0   10.0  
    \arrow <12pt> [.1,.4] from    10.0   -10.0  to -5.0   5.0  

    \arrow <12pt> [.1,.4] from   9.0   6.0  to   3.0  0.0  %
    \arrow <12pt> [.1,.4] from  13.5   1.5  to  10.5 -1.5  %
    \arrow <12pt> [.1,.4] from   9.0   6.0  to  13.5  1.5  %
    \arrow <12pt> [.1,.4] from   3.0   0.0  to   7.5 -4.5  %

    \setdashes

    \plot -5.0   -8.0   10.0  7.0   /  %
    \plot  0.0  -12.0   14.0  2.0   /  %


    \setsolid

    \put {\large u} at -3.5 4.0
    \put {\large v} at  8.5 9.0
    \put {$\bullet$}  at  0.0 0.0  

    \put {$ \gamma u = -1 $} at -5.0 -8.3 %
    \put {$ \gamma u = -2 $} at 0.0 -12.3 %

    \put {\bf P}  at 9.0 6.4  %
    \put {\bf P'} at 2.7 0.3  %
    \put {\bf Q}  at 13.9 1.3  %
    \put {\bf Q'} at 11.0 -1.9 %

    \put {$ {\bf \large e}_0 ({\bf P})  $} at 12.0 4.0  %
    \put {$ {\bf \large e}_0 ({\bf P'}) $} at 6.0 -1.9  %

    \put {$ {\bf \large e}_1 ({\bf Q}) $}  at 12.5 -0.5  %
    \put {$ {\bf \large e}_1 ({\bf P}) $} at  5.0  3.0  %

\endpicture}
$$
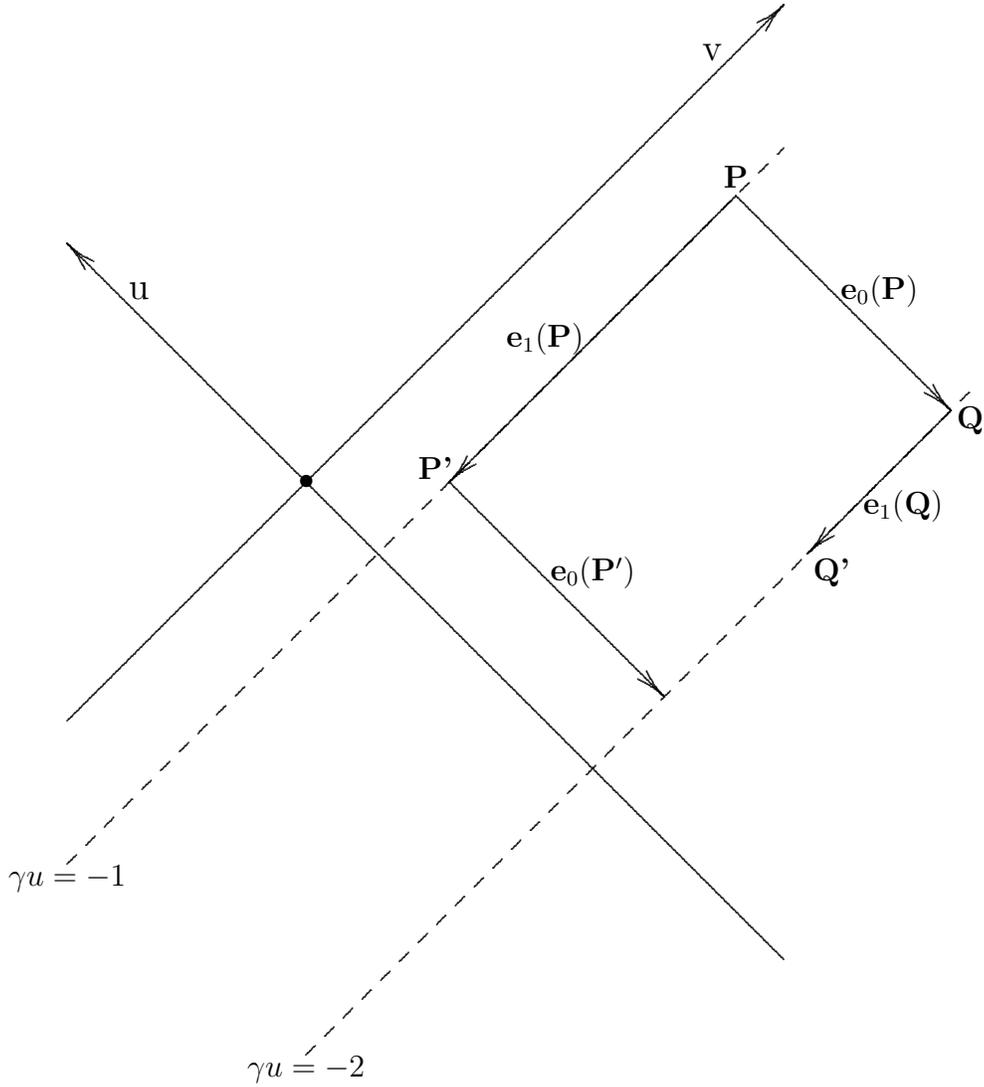
{\centerline {\bf Figure 4: Cartan's Torsion}} \newline


\section{The Gravitational Frequency Shift}
\setcounter{equation}{0}

For a first orientation,
we draw the two hyperbolic worldlines of two observers
who experience a homogeneous gravitational field.
[{\bf See Figure~5.}]
Their 4-velocities ${\bf V}$ are given by Eq.(\ref{eq:MinkVvector}) for negative $u$
\begin{equation} \label{eq:Vvector}
{\bf V} \, = \, V^{\mu} \, {\bf e}_{\mu} \, = \,
- \, V^{0} \, \gamma \, u \, \partial_{u} \, - \,
\frac{1}{2\, V^{0} \, \gamma \, u} \, \partial_{v} \, .
\end{equation}
$$
\hskip 0.5 in
{\beginpicture
    \setcoordinatesystem units <.25in,.25in> point at 0 0

    \arrow <12pt> [.1,.4] from  -5.0    0.0   to  13.0   0.0  
    \arrow <12pt> [.1,.4] from   0.0  -10.0   to   0.0  11.0  

    \arrow <12pt> [.1,.4] from    -5.0    -5.0  to  10.0   10.0  
    \arrow <12pt> [.1,.4] from    10.0   -10.0  to  -5.0    5.0  

    \arrow <12pt> [.1,.4] from    0.0  -6.0  to  0.0  -3.0  %
    \plot  0.02 -6.0   0.02 -3.0 /  %
    \plot -0.02 -6.0  -0.02 -3.0 /  %

    \arrow <12pt> [.1,.4] from    6.0  0.0  to  6.0  3.0  %
    \plot  6.02 0.0   6.02 3.0 /  %
    \plot  5.98 0.0   5.98 3.0 /  %

    \arrow <12pt> [.1,.4] from    4.50   3.88  to  7.0  6.66  
    \plot 4.50 3.86  7.00 6.64 /  %
    \plot 4.50 3.84  7.00 6.64 /  %
    \plot 4.50 3.90  7.00 6.68 /  %
    \plot 4.50 3.92  7.00 6.68 /  %

    \setquadratic

    \plot  10.50  -9.50    6.00  0.00    10.50   9.50  /  
    \plot  10.53  -9.50    6.03  0.00    10.53   9.50  /  

    \plot   3.10 -14.00   0.05  -5.20    4.50   3.90  /  
    \plot   3.13 -14.00   0.08  -5.20    4.53   3.90  /  

    \setlinear
    \setdashes
 
    \plot  0.0   -6.00   10.0   4.00   /  
    \plot 10.0   -3.96   -5.0  11.04   /  
    \plot  2.05 -14.06   -5.0  -7.00   /  

    \setsolid

    \arrow <12pt> [.1,.4] from    9.0  3.0   to  10.0   4.0   
    \arrow <12pt> [.1,.4] from   -4.0 10.04  to  -5.0  11.04  

    \put {\large u} at -3.5 4.0
    \put {\large v} at  8.5 9.0
    \put {$\bullet$}  at 0 0  
    \put {\large t} at -0.3 10.0
    \put {\large x} at  12.0  0.3

    \put {higher}   at 11.0 -8.0
    \put {observer} at 11.0 -8.4

    \put {lower}    at 2.5 -10.4
    \put {observer} at 2.5 -10.8

    \put {descending} at -3.0 11.0
    \put {photon}     at -3.0 10.6

    \put {rising} at 11.0 4.0
    \put {photon} at 11.0 3.6

\endpicture}
$$
\vskip 1.0in
{\centerline {\bf Figure 5: The Gravitational Frequency Shift}}
\medskip
\noindent
\hrulefill
\medskip
\newline

Since $ V^{0} $ is constant, the vectors are parallel
in the usual affine sense along lines $ u = constant $.
These are the null lines of rising photons emitted
when the lower observer has precisely the same velocity
as the higher observer at the reception of the light.
There is no frequency shift for a rising photon.
It is obvious from the figure that a falling photon is perceived
by the lower observer to originate from an approaching source
and thus be blue-shifted.
To calculate the shift, we remember that the ratio of the frequency $ \nu $
of the emitter to the frequency $ \nu' $ of the receiver is given by 
\begin{equation}
\frac{\nu'}{\nu} \, = \, \frac{{\bf k}\cdot{\bf V'}}{{\bf k}\cdot{\bf V}} \, .
\end{equation}
Here, $ {\bf V} $ and $ {\bf V'} $ are the 4-velocities of
emitter and receiver, respectively,
while $ {\bf k} $ is the 4-vector of the photon.

For the falling photon the $ {\bf k} $ null vector
has a $u$-component only, say, 
\begin{equation}
{\bf k} \, \sim \, ( \, 1 , \, 0 \, ) \, .
\end{equation}
This is given, with Eq.(\ref{eq:Vvector}),
\begin{equation}
\frac{\nu'}{\nu} \, = \, \frac{-1}{2\, V^{0} \, \gamma \, u'}
\, : \,\frac{-1}{2\, V^{0} \, \gamma \, u} \, = \,
\left \vert \, \frac{u}{u'} \, \right \vert \, .
\end{equation}
Since $ \vert \, u' \, \vert $ is smaller than $ \vert \, u \, \vert $,
we obtain a blue-shift.
We now also see that this blue-shift for a falling photon
is the physical equivalent of the non-closure of our square $PQP'Q'$.
The ratio in length of $ PP' / QQ' $ is the blue-shift $ \nu' / \nu $.

As we had seen in Eq.(\ref{eq:v-v0}),
two different observers differ only in their value of the parameter $ v_{0} $
\begin{equation}
u \, = \, - \, \frac{1}{ 2 \, ( \, v \, - \, v_{0} \, ) \, ( \, V^{0} \, \gamma \, )^{2} } \, .
\end{equation}
This shows that the ratios of their $u$-values, $ \vert \, u / u' \, \vert $,
goes towards $1$ for  $ v \rightarrow \infty $.
The blue-shift vanishes asymptotically.

\section{Fermi's Torsion}
\setcounter{equation}{0}

Almost simultaneously with \'Elie Cartan's 1922 introduction of torsion,
this concept and its importance for physics was also discovered
by a student at Italy's famous Scuola Normale Superiore in Pisa.
Enrico Fermi, then aged 21, had sent a note to the
Rendiconti of the Accademia dei Lincei on
\lq\lq {\it Sopra i fenomeni che avvengono in vicinanza di una linea oraria}\rq\rq\
(On the phenomena that occur in the neighborhood of a time-like worldline).\cite{Fermi}
The short paper, that appeared in three pieces, introduced
what became known as Fermi transport of vectors.\cite{Walker}

Let $ {\bf U} $ be a time-like unit vector field tangent to
a bunch of worldlines.
We have
\begin{equation} \label{eq:uu}
{\bf U} \cdot {\bf U} \, = \, 1 \, .
\end{equation}
The acceleration along the worldlines is given by
\begin{equation}
{\bf A} \, = \, {\bf \dot{U}} \, \equiv \, \frac{D{\bf U}}{Ds} \, ,
\end{equation}
where $ {\bf A} $ is the acceleration vector and $s$
measures proper time along the worldlines.
Clearly, because of Eq.(\ref{eq:uu}),
\begin{equation}
{\bf \dot{U}} \cdot {\bf U} \, = \, {\bf A} \cdot {\bf U} \, = \, 0 \, ,
\end{equation}
the acceleration vector is orthogonal to the four-velocity ${\bf U}$.
The two vectors ${\bf U}$ and $ {\bf A} $ form a bivector
$ {\bf U} \wedge {\bf A} $ that describes a Lorentz boost
for the frame that has the four-velocity vector ${\bf U}$.

For a vector ${\bf V}$ in the frame with four-velocity
\begin{equation}
{\bf U} \, = \, {\bf e}_{0} \, ,
\end{equation}
the time-part ${\bf V_{\Vert}}$ is given by the projection
of ${\bf V}$ on ${\bf U}$
\begin{equation}
{\bf V_{\Vert}}
\, \equiv \,
( {\bf V} \cdot {\bf U} ) \, {\bf U}
\, = \,
V^{0} \, {\bf e}_{0} \, ,
\end{equation}
while the space-part ${\bf V_{\perp}}$ is defined as
\begin{equation}
{\bf V_{\perp}}
\, \equiv \,
V^{j} \, {\bf e}_{j}
\, = \,
V^{\mu} \, {\bf e}_{\mu} \, - \, V^{0} \, {\bf e}_{0}
\, = \,
{\bf V} \, - \, ( {\bf V} \cdot {\bf U} ) \, {\bf U} \, .
\end{equation}
Fermi transport for a vector ${\bf V}$ along a worldline with
unit tangent ${\bf U}$ is characterized by a boost that compensates
the acceleration $ {\bf A} $.
The time-part of ${\bf V}$ remains unchanged,
while the space-part becomes subject to a rotation
preserving the length of the space-part of the vector ${\bf V}$ and,
thus, its four-dimensional length.
The Fermi connection is thus a metrical connection.

As an example of Fermi transport,
we return to Born motion in the Minkowski plane,
as described in {\bf Section \ref{sect:BM}},
with the use of \lq\lq polar coordinates\rq\rq.
(They are called \lq\lq Rindler coordinates\rq\rq\ by some authors.)
[{\bf See Figure~6.}]

The unit tangent vector, $ {\bf U} $, along the hyperbolae
will remain a unit tangent vector along the hyperbolae under
Fermi transport.
Along the radial geodesics, however, Fermi transport becomes
Levi-Civita's parallel transport.
It is clear then that we have torsion since parallelograms do not close.
It is easy to see that the torsion becomes
the cause of frequency shifts of light.
We have the metric Eq.(\ref{eq:BMpolar})
\begin{equation}
ds^{2} \, = \, \rho^{2} \, d\tau^{2} \, - \, d\rho^{2} \, .
\end{equation}
Null lines are given by $ ds = 0 $ and thus
\begin{equation}
\pm \, d\tau \, = \, \frac{d\rho}{\rho} \, .
\end{equation}
A light ray emitted from $ \rho_{1} $ at time $ \tau_{1} $ will be
received at $ \rho_{2} $ at time $ \tau_{2} $:
\begin{equation}
\pm \, \int_{\tau_{1} + d\tau_{1}}^{\tau_{2} + d\tau_{2}} d\tau
\, = \,
\pm \, \int_{\tau_{1}}^{\tau_{2}} d\tau
\, = \,
\int_{\rho_{1}}^{\rho_{2}} \frac{d\rho}{\rho}
\, = \,
\ln \frac{\rho_{2}}{\rho_{1}} \, .
\end{equation}

$$
\hskip 0.25 in
{\beginpicture
    \setcoordinatesystem units <.25in,.25in> point at 0 0

    \arrow <12pt> [.1,.4] from  -1.0    0.0   to  15.0   0.0  
    \arrow <12pt> [.1,.4] from   0.0  -10.0   to   0.0  11.0  

    \arrow <12pt> [.1,.4] from   -1.0   -1.0  to  10.0   10.0  
    \arrow <12pt> [.1,.4] from    10.0   -10.0  to -1.0   1.0  

    \setquadratic

\plot   9.000 -8.062
        8.000 -6.928
        7.000 -5.744
        6.500 -5.123
        6.000 -4.472
        5.000 -3.000
        4.500 -2.061
        4.000  0.000
        4.500 +2.061
        5.000 +3.000
        6.000 +4.472
        6.500 +5.123
        7.000 +5.744
        8.000 +6.928
        9.000 +8.062
/

\plot   15.000 -11.180   
        14.000  -9.797
        13.500  -9.069
        13.000  -8.306
        12.000  -6.633
        11.500  -5.678
        11.000  -4.582
        10.000   0.000
        11.000  +4.582
        11.500  +5.678
        12.000  +6.633
        13.000  +8.306
        13.500  +9.069
        14.000  +9.797
        15.000 +11.180
/  

    \setlinear

     \arrow <12pt> [.1,.4] from   4.0  0.0   to   5.50  0.00  
     \plot 4.00  0.02   5.50  0.02   /  
     \plot 4.00 -0.02   5.50 -0.02   /  

     \arrow <12pt> [.1,.4] from   4.0  0.0   to   4.00  1.40  
     \plot 3.98  0.00   3.98  1.40   /  
     \plot 4.02  0.00   4.02  1.40   /  

     \arrow <12pt> [.1,.4] from   10.00  0.00   to   11.5  0.00  
     \plot 10.00  0.02   11.5  0.02   /  
     \plot 10.00 -0.02   11.5 -0.02   /  

     \arrow <12pt> [.1,.4] from   10.0  0.0   to   10.0  1.40  
     \plot  9.98  0.00    9.98  1.40   /  
     \plot 10.02  0.00   10.02  1.40   /  



     \put {$ \odot $}  at    4.000   0.000
     \put {$ \odot $}  at   10.000   0.000
     \put {$ \odot $}  at    5.800   4.190
     \put {$ \odot $}  at   14.500  10.500


\setdashes
     \plot  4.000   0.000   15.000  11.00   /  
     \plot 10.000   0.000    5.800  4.190   /  
\setsolid

     \plot  5.800   4.190   14.500  10.500   /  

    \put {\large u} at -1.5  1.0
    \put {\large v} at  8.5  9.0
    \put {$\bullet$}  at  0 0       
    \put {\large t} at -0.3 10.0
    \put {\large x} at 14.0  0.3

\endpicture}
$$
\vskip 2.7in
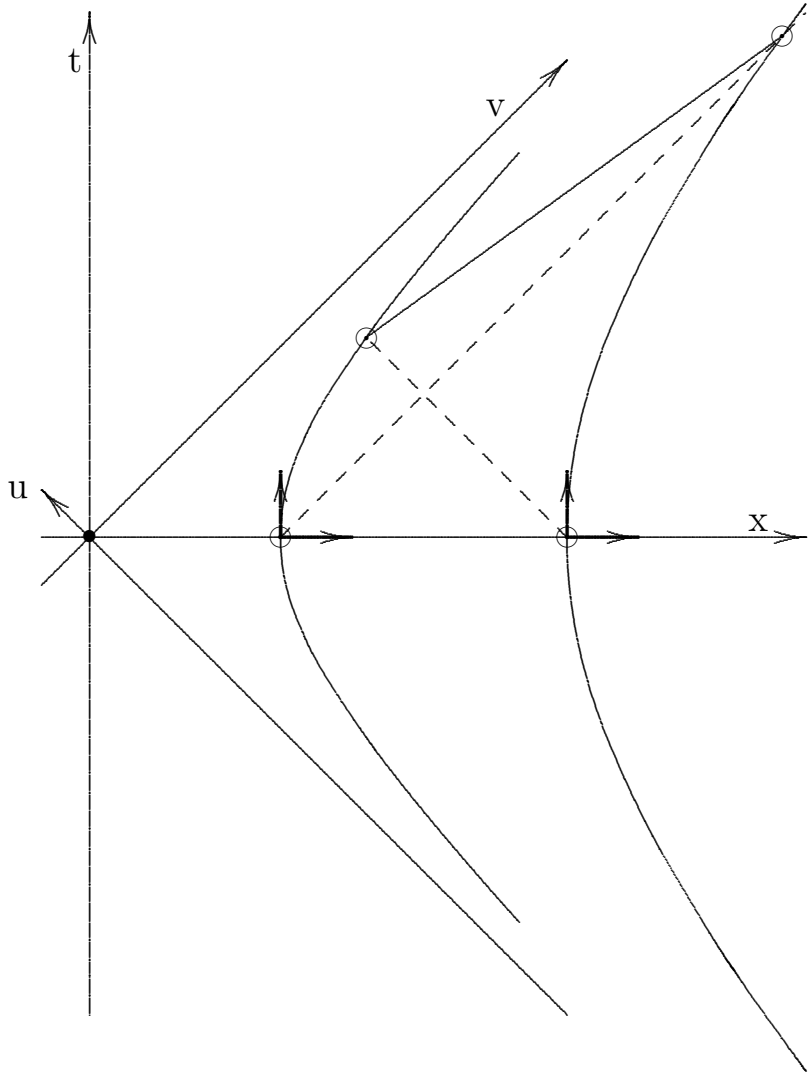
{\centerline {\bf Figure 6: Fermi's Torsion}}
\medskip
\noindent
\hrulefill
\medskip
\newline

\noindent
A second light ray emitted at $ \tau_{1} + d\tau_{1} $,
will be received at $ \tau_{2} + d\tau_{2} $.
It follows that
\begin{equation}
d\tau_{1} \, = \, d\tau_{2} \, .
\end{equation}
Since frequencies go inversely to proper times,
$ds$, we have the gravitational shift 
\begin{equation}
\frac{\nu_{1}}{\nu_{2}}
\, = \,
\frac{ds_{2}}{ds_{1}}
\, = \,
\frac{\rho_{2} \, d\tau_{2}}{\rho_{1} \, d\tau_{1}}
\, = \,
\frac{\rho_{2}}{\rho_{1}} \, .
\end{equation}
Going back to Eq.(\ref{eq:BMvectors}),
we can now calculate the torsion.
We have
\begin{equation}
{\mbox {\boldmath $\omega$}}^{0} \, = \, \rho \, {\bf d}{\mbox {\boldmath $\tau$}} \, , \qquad
{\mbox {\boldmath $\omega$}}^{1} \, = \, {\bf d}{\mbox {\boldmath $\rho$}} \, ,
\end{equation}
and
\begin{equation}
ds^{2} \, = \,
\left[ \, {\mbox {\boldmath $\omega$}}^{0} \right]^{2}
\, - \, \left[ \, {\mbox {\boldmath $\omega$}}^{1} \right]^{2} \, .
\end{equation}
The connection form is zero since vectors do not change their
components under Fermi transport in the Minkowski plane.
We thus have for the torsion form's vector components, ${\bf \Theta}^{\mu}$,
\begin{equation}
{\bf \Theta}^{0} \, = \, {\bf d}{\mbox {\boldmath $\omega$}}^{0}
\, = \, {\bf d}{\mbox {\boldmath $\rho$}} \wedge {\bf d}{\mbox {\boldmath $\tau$}}
\, = \, \frac{1}{\rho}
     \, {\mbox {\boldmath $\omega$}}^{1} \wedge {\mbox {\boldmath $\omega$}}^{0}
\, , \qquad
{\bf \Theta}^{1} \, = \, {\bf d}{\mbox {\boldmath $\omega$}}^{1} \, = \, 0 \, .
\end{equation}
With
\begin{equation}
{\mbox {\boldmath $\Theta$}}^{\mu}
\, = \,
\frac{1}{2} \, T^{\mu}{}_{\lambda\nu} \
{\mbox {\boldmath $\omega$}}^{\lambda} \wedge {\mbox {\boldmath $\omega$}}^{\nu} \, ,
\end{equation}
we obtain
\begin{equation}
T^{0}{}_{01} \, = \, - \, \frac{1}{\rho}
\, , \qquad
T^{1}{}_{01} \, = \, 0 \, .
\end{equation}
We can express the gravitational shift in terms of torsion coefficients.

\section{Reference Frames}
\setcounter{equation}{0}

Enrico Fermi had introduced his vector transport in
approaching a physical description of a gravitational field.
It would seem natural for a physicist to interpret
the hyperbolae of Born motion as the worldlines
of rigid flat slabs made of incompressible matter.
But all attempts to introduce rigid bodies into relativity
came to naught.
This became particularly clear through a one-page paper by
Pawel Sigmundovich Ehrenfest in Leiden
that was the death knell for rigidity and led to the introduction
of non-Euclidean geometry into relativity theory.
The paper posing Ehrenfest's paradox pointed out that
a circle of radius $R$ centered on the axis of rotation
in a rigid body and orthogonal to it would not show
Lorentz's contraction of its radius while its circumference would.
The deviation from $2\pi$ in the ratio of circumference to radius
of the circle would thus indicate the presence of non-Euclidean geometry.
 
We shall define our reference system abstractly as an orthonormal frame
in every point of the spacetime manifold.
Nowadays one calls this a section of the frame bundle.
How this can be physically realized is another question.
Mathematically this section is given by 
\begin{equation}
{\bf e}_{\nu}(x^{\lambda}) \, .
\end{equation}

When Einstein wrote about accelerated reference systems,
it was assumed that meter sticks and clocks in such systems
would be unchanged.
Experiments done with muons in synchrocyclotrons have confirmed
this assumption to high accuracy.
Since clocks and meter sticks can be represented as vectors
in four dimensions, it follows that the frames themselves,
their metric, and, thus,
their orthonormality is not affected by acceleration.

The acceleration of observers or,
as Einstein thought equivalent to it,
a gravitational field
are not due to the curvature of the space-time manifold.
The falling of apples from Earth-bound trees and
the weight we are experiencing while standing on the surface
of this planet is the result of an acceleration
of our reference frame that can be described
mathematically as the torsion field of Minkowski spacetime.
The presence of curvature near the surface of the Earth,
that is, a non-vanishing Riemann tensor, corresponds to
second order derivatives of the Newtonian potential and is a
small higher order effect.
The mantra \lq\lq gravitation is curvature\rq\rq\ is misleading
if one associates gravitation with the falling of apples.

\section{The Meaning of Einstein's First Principle of Equivalence}
\setcounter{equation}{0}

What we now call the principle of equivalence,
to be more precise, the weak principle, says:
\begin{quote}
\lq\lq The worldline of a spinless test particle,
moving under the influence of gravitational fields only,
depends on its initial position and velocity,
but not its mass and composition.\rq\rq
\end{quote}
Apart from its fanciful spacetime formulation,
Isaac Newton could have put that into his
\lq\lq {\it Principia Philosophiae Naturalis}\,\rq\rq.
What happened to the equivalence of acceleration and gravitation?

For a Riemannian spacetime, using \'Elie Cartan's first fundamental equation,
we can formulate a kind of equivalence of inertial and
gravitational acceleration.
The equation
\begin{equation}
{\bf \Theta}^{\mu} \, = \,
{\bf d}{\mbox {\boldmath $\omega$}}^{\mu}
\, - \,
{\mbox {\boldmath $\omega$}}^{\mu}{}_{\nu} \wedge {\mbox {\boldmath $\omega$}}^{\nu}
\end{equation}
can be written down twice: for the same frame
$ {\mbox {\boldmath $\omega$}}^{\nu} $ and as
\begin{equation}
{\bf \Theta}'^{\mu} \, = \,
{\bf d}{\mbox {\boldmath $\omega$}}^{\mu} \, - \,
{\mbox {\boldmath $\omega$}}'^{\mu}{}_{\nu} \wedge {\mbox {\boldmath $\omega$}}^{\nu}
\end{equation}
If we put the torsion, $ {\bf \Theta}^{\mu} $, equal to zero in the first equation,
we have
\begin{equation}
{\bf d}{\mbox {\boldmath $\omega$}}^{\mu} \, = \,
{\mbox {\boldmath $\omega$}}^{\mu}{}_{\nu} \wedge {\mbox {\boldmath $\omega$}}^{\nu}
\end{equation}
where the connection forms, $ {\mbox {\boldmath $\omega$}}^{\mu}{}_{\nu} $,
describe a gravitational field in a spacetime without torsion.
If we put the connection forms, $ {\mbox {\boldmath $\omega$}}'^{\mu}{}_{\nu} $,
in the second equation equal to zero, we have
\begin{equation}
{\bf d}{\mbox {\boldmath $\omega$}}^{\mu} \, = \, {\bf \Theta}'^{\mu}
\end{equation}
and thus
\begin{equation}
{\bf \Theta}'^{\mu} \, = \,
{\mbox {\boldmath $\omega$}}^{\mu}{}_{\nu} \wedge {\mbox {\boldmath $\omega$}}^{\nu} \, .
\end{equation}
The vanishing connection forms $ {\mbox {\boldmath $\omega$}}'^{\mu}{}_{\nu} $ imply
teleparallelism in spacetime.
The tensor character of the torsion follows from the fact that
\begin{equation}
{\bf \Theta}'^{\mu} \, - \, {\bf \Theta}^{\mu}
\, = \,
{\bf \Theta}'^{\mu}
\, = \,
- \, \left( \, {\mbox {\boldmath $\omega$}}'^{\mu}{}_{\nu} \, - \,
{\mbox {\boldmath $\omega$}}^{\mu}{}_{\nu} \, \right) \wedge {\mbox {\boldmath $\omega$}}^{\nu}
\, = \,
{\mbox {\boldmath $\omega$}}^{\mu}{}_{\nu} \wedge {\mbox {\boldmath $\omega$}}^{\nu} \, ,
\end{equation}
is obtained through the difference of two connection forms.
The justification for the choice of a vanishing connection
in the accelerated frame is suggested by the
way in which measurements are carried out.

Encouraged by Ernst Mach's critique of Isaac Newton's mechanics,
Einstein tried to argue that the equivalence of inertia and
gravitation made acceleration against absolute space obsolete.
Only relative acceleration of bodies was supposed to be observable.
All inertial forces could then also be interpreted as
gravitational ones.
However, these gravitational fields had no sources and were
generated by coordinate transformations.

For the modern physicist,
acceleration against Newton's absolute space is no longer
so implausible as it appeared a century ago.
The vacuum of spacetime is filled with the fluctuations of
all fields and houses a non-zero Higgs field and a metric too.
Absolute space is realized by a local inertial system.

The modern view of Einstein's theory of gravitation was stated
by Sir Hermann Bondi at the occasion of the
Centenary of Einstein's birthday in 1979.
He wrote:
\begin{quote}
\lq\lq From this point of view,
Einstein's elevators have nothing to do with gravitation;
they simply analyze inertia in a perfectly Newtonian way.
Thus, the notion of general relativity does not in fact
introduce any post-Newtonian physics;
it simply deals with coordinate transformations.
Such a formalism may have some convenience,
but physically it is wholly irrelevant.
It is rather late to change the name of Einstein's theory
of gravitation, but general relativity is a physically
meaningless phrase that can only be viewed as a historical
memento of a curious philosophical observation.\rq\rq\cite{Bondi1979}
\end{quote}
Physically, it is a clearcut case whether the accelerometer
of an observer shows zero or not; this holds too for the
mathematics.
Whether two events in Minkowski spacetime are connected by a
straight timelike line or by a hyperbolic path can easily be
distinguished because the hyperbolic path has a smaller proper time.
So it is clear that the accelerated frame has to be described
as the frame that has torsion and there is nothing left of
the so-called equivalence or relativity.

What has been gained, however, is an extension of the class
of reference systems from the inertial frames of Minkowski
spacetime to all smooth sections of the frame bundle.

This does not lead to the most general kind of torsion since
\begin{equation}
{\bf \Theta}'^{\mu} \, = \, {\bf d}{\mbox {\boldmath $\omega$}}^{\mu}
\end{equation}
gives that
\begin{equation}
{\bf d}{\bf \Theta}'^{\mu} \, = \, 0 \, .
\end{equation}
Because the curvature, $ {\mbox {\boldmath $\Omega$}}'^{\mu}{}_{\nu} $, of the
connection, $ {\mbox {\boldmath $\omega$}}'^{\mu}{}_{\nu} $, vanishes,
the integrability conditions of the torsion
\begin{equation}
{\mbox {\boldmath $\nabla$}} {\bf \Theta}'^{\mu}
\, = \,
{\bf d}{\bf \Theta}'^{\mu} \, - \,
{\mbox {\boldmath $\omega$}}'^{\mu}{}_{\nu} \wedge {\bf \Theta}^{\nu}
\, = \,
- \, {\mbox {\boldmath $\Omega$}}'^{\mu}{}_{\nu} \wedge {\mbox {\boldmath $\omega$}}^{\nu}
\end{equation}
give also that
\begin{equation}
{\mbox {\boldmath $\nabla$}} {\bf \Theta}'^{\mu} \, = \, 0 \, .
\end{equation}
It is a remarkable fact that the frames expressing linear
and rotational acceleration can be interpreted---via torsion---as
an invariant property of spacetime.

\section{The Use of Frames}
\setcounter{equation}{0}

Gregorio Ricci-Curbastro used frames in his calculus and so did the
French geometer Gaston Darboux.
His technique of moving frames was developed to perfection
by his student \'Elie Cartan.
Students of general relativity followed Einstein in the
unphysical use of coordinate bases.
In 1929, Eugene Wigner, Hermann Weyl, and Vladimir Fock
introduced orthonormal frames into general relativity to
formulate the Dirac equation for curved spacetimes.

While frames are important for the introduction of a spin structure
and the four-dimensional Gauss-Bonnet theorem,
there are critical physical reasons for their use.
Since physics generally deals with spatially extended objects
(that is, fields) by working with their frame components,
we can also consider global Lorentz transformations for a
section of the frame bundle.

The most important aspect of the frame is that we can consider
it as a simplified geometric representation of the quantized
measuring instruments
(Bohr radius of the hydrogen atom, Cesium hyperfine structure
line defining the second, etc.)
that determine the metric.
Mathematicians often seem to ignore that all measurements are relative.

\section{Einstein's Torsion}
\setcounter{equation}{0}

Torsion, invented by \'Elie Cartan and announced in a three page
note to the French Academie, was introduced by Emile Borel in
the session of February 27, 1922.
Its publication in the Comptes Rendus bore the title
\lq\lq {\it Sur une g\'en\'eralisation de la notion de courbure de
Riemann et les espaces \`a torsion}\rq\rq.
Einstein learned about this new revolutionary concept in
geometry only four weeks later.
His friend Paul Langevin had invited him to give a lecture
at the Coll\`ege de France in Paris on March 31, 1922.

In the aftermath of World War I, the first lecture by a professor
from the archenemy country was a highly charged political affair.
To cut down on demonstrations,
it was by invitation only and the French Prime Minister, Paul Painlev\'e,
stood at the door checking the invitations.
During this lecture week, Jacques Hadamard, professor at the
Coll\`ege de France, gave a party for Einstein.
Among his guests was \'Elie Cartan who was to meet Einstein there.
Cartan thought torsion might have important physical applications
and used the occasion to tell Einstein about his recent discovery.
He tried to explain the novel concept to him using the simplest
example well known to mapmakers and navigators.
[{\bf See Figure~7.}]

$$
\vspace{5.5in}
{\beginpicture
    \setcoordinatesystem units <1.0in,1.0in> point at 0.0 2.5

    \circulararc                     360 degrees from  2.5    0.00   center at  0.0  0.0  

    \ellipticalarc axes ratio 4:1   -180 degrees from  2.5    0.00   center at  0.0  0.0  
    \ellipticalarc axes ratio 4.3:1 -120 degrees from  1.5    2.00   center at  0.0  2.2  
    \ellipticalarc axes ratio 4.3:1 -195 degrees from  2.2    1.18   center at  0.0  1.1  
    \ellipticalarc axes ratio 2.9:1 -160 degrees from  2.191 -1.20   center at  0.0 -1.1  

    \ellipticalarc axes ratio 1:5     180 degrees from  0.0   -2.50   center at  0.0  0.0  %
    \ellipticalarc axes ratio 1:1.5   180 degrees from  0.0   -2.50   center at  0.0  0.0  %
    \ellipticalarc axes ratio 1:1.5  -180 degrees from  0.0   -2.50   center at  0.0  0.0  %

    \arrow <12pt> [.1,.4] from  -1.092 1.887  to -0.400  2.100  
    \arrow <12pt> [.1,.4] from   0.344 1.805  to  0.900  2.150  

    \arrow <12pt> [.1,.4] from  -1.595 0.740  to -0.900  0.960  
    \arrow <12pt> [.1,.4] from   0.485 0.595  to  1.100  0.970  
    \arrow <12pt> [.1,.4] from   1.590 0.740  to  2.050  1.230  

    \arrow <12pt> [.1,.4] from  -1.638 -0.472  to -0.950  -0.250  
    \arrow <12pt> [.1,.4] from   0.485 -0.615  to  1.05  -0.250  
    \arrow <12pt> [.1,.4] from   1.636 -0.473  to  2.1500  0.0  

    \arrow <12pt> [.1,.4] from  -1.197 -1.740  to -0.5 -1.5  
    \arrow <12pt> [.1,.4] from   0.335 -1.853  to  0.950 -1.5  
    \arrow <12pt> [.1,.4] from   1.195 -1.739  to  1.750 -1.3  


\setquadratic

    \plot   -0.64  2.02     -0.580  1.93      -0.62  1.825  /  
    \plot    0.69  2.01      0.745  1.93       0.73  1.833  /  

    \plot   -1.14  0.88     -1.07  0.77      -1.12  0.65  /  
    \plot    0.89  0.84      0.94  0.74       0.91  0.63  /  
    \plot    1.87  1.03      1.93  0.95       1.93  0.84  /  

    \plot   -1.20 -0.55     -1.15 -0.44      -1.20  -0.34  /  
    \plot    0.86 -0.58      0.89 -0.48       0.84  -0.38  /  
    \plot    2.01 -0.37      2.03 -0.26       1.97  -0.17  /  

    \plot   -0.75  -1.59     -0.7  -1.7       -0.75  -1.82  /  
    \plot    0.75  -1.81      0.77 -1.72       0.73  -1.63  /  
    \plot    1.6   -1.63      1.61 -1.53       1.56  -1.45  /  

\setlinear


    \plot -2.2200 0.1000   -0.900 -0.200  /  
    \plot -2.2200 0.1000   -2.1200 -0.70  /  
    \plot -0.900 -0.200   -0.800 -0.990  /  
    \plot -2.1200 -0.70   -0.800 -0.990  /  

\endpicture}
$$
\medskip
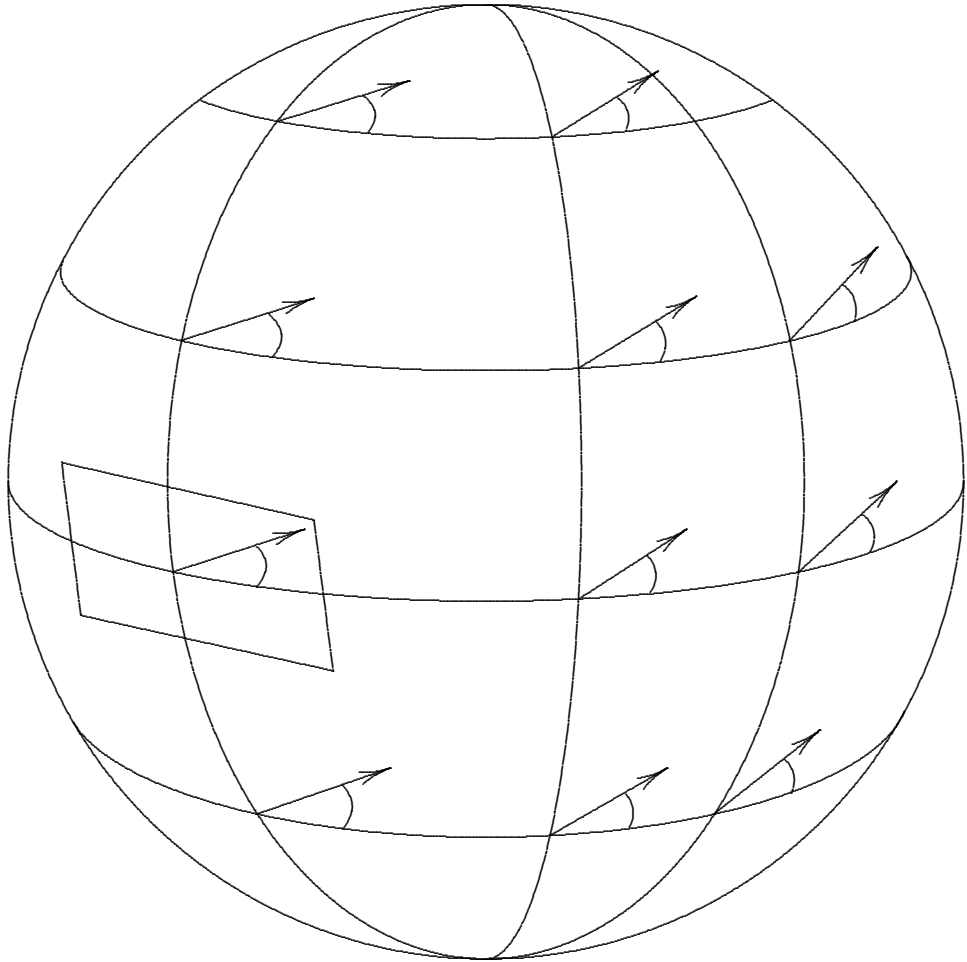
{\centerline {\bf Figure 7: Einstein's Torsion}} \newline
Cartan's example of a connection with torsion that confused Einstein.
The arrows are the same length and make constant angles with their respective latitude lines.
They lie in tangent planes located at their origins.
\newline


If one takes a sphere, removes the North and South poles,
and puts unit vectors pointing North along meridians 
and along latitude circles pointing East,
one has created an orthonormal frame everywhere.
If one now moves vectors tangent to the sphere by keeping
their frame components constant,
one has a connection with torsion.
The lines that have such teleparallel vectors as tangents
have been long known as loxodromes or rhumb lines,
discovered in 1537 as \lq\lq {\it curvas dos rumos}\rq\rq\
by a {\it marra\~no}, the Portuguese geometer Pedro Nu\~nez.
Cartan's example of torsion was the exact analog of Fermi
transport for a two dimensional spacetime that had been
published on January 22, 1922, a month before Cartan.
It was uncanny that seven years later Einstein wrote to Cartan:
\begin{quote}
\lq\lq I didn't at all understand the explanations you gave me in Paris;
still less was it clear to me how they might be made useful
for physical theory.\rq\rq\cite{Einstein1929}
\end{quote}
The occasion of this remark was a letter by Cartan pointing
out that Einstein's teleparallelism was a special case of Cartan's torsion.
But since publication of his first paper on teleparallelism
in the year before, Einstein had learned of Eisenhart's work
and that of Roland Weitzenb\"ock.
In fact, Weitzenb\"ock had given a supposedly complete bibliography
of papers on torsion with fourteen references without mentioning
\'Elie Cartan.
Since Weitzenb\"ock, of Amsterdam University in the Netherlands,
had written the review article on Differential Invariants
for the Encyclopedia of Mathematical Science,
he certainly would have been aware of Cartan's activity
in this field.
A bizarre circumstance suggested that this omission may
have been deliberate:
In 1923 Weitzenb\"ock published a modern monograph on the
Theory of Invariants that included tensor calculus.
In the innocent looking Preface, we find that
the first letter of the first word in each of its twenty-one sentences
spell out:
\begin{quote}
\lq\lq NIEDER MIT DEN FRANZOSEN\rq\rq\ (Down with the French!)
\end{quote}
Weitzenb\"ock, a Prussian-born army officer of World War I,
had in fact mentioned his own kind of torsion as a possible
connection in the review article of the Encyclopedia in 1923.
But his silence on the French work on torsion in his 1928
review appeared to be a subliminal intellectual continuation of warfare.
To set the record straight, Einstein asked Cartan to describe
the history of torsion in an article in the leading German
mathematics journal {\it Mathematische Annalen} as
\lq\lq {\it Notice historique sur la notion de parall\'elisme absolu}\rq\rq\cite{Cartan1930}.

Einstein had introduced frames for a generalized
theory using teleparallelism in 1928.
In fact, Wigner's paper was based on Einstein's tetrad formalism.
Einstein was using torsion for the electromagnetic potentials
in his unified field theory.
The resulting opus provoked Wolfgang Pauli's scathing critique
in a letter to Einstein of December 19, 1929.
Concluding his letter, Pauli wrote:
\begin{quote}
\lq\lq I am also not so naif as to believe you would change
your mind because of criticism by others.
But I would bet you any odds that in a year at the latest
you will have dropped the whole teleparallelism
as you earlier dropped your affine theory.
And I do not wish to provoke your spirit of contradiction by
continuing this letter to not delay the approaching natural
demise of the theory of teleparallelism.
In this definite hope of winning the bet,
I wish you a merry Christmas affectionately, truly yours, W.Pauli\rq\rq\cite{Pauli}
\end{quote}
Einstein did not accept Pauli's criticism or his bet.
It took him two year's to drop teleparallelism and to come up with
a new variant of Kaluza's five-dimensional theory.

\section{The Notion of Torsion}
\setcounter{equation}{0}

Looking back, we may wonder why the concept of torsion and
its importance for physics were apparently so difficult to grasp.
There are, in fact, more examples in the literature.
The gravitational \lq\lq {\it Handbuch}\,\rq\rq\ by Charles Misner,
Kip Thorne, and John Archibald Wheeler deals with the subject
in their {\bf Box 10.20} where the torsion is set to zero by a circular argument \cite{MTW}.
This was discovered by Borut Gogala \cite{Gogala} and James Nester \cite{Nester}.
But then even expert differential geometers could have trouble with the concept:
in his \lq\lq {\it Notes on Differential Geometry},\,\rq\rq\ Noel J. Hicks declares
\begin{quote}
\lq\lq As far as we know, there is no motivation for the word \lq torsion\rq\
to describe the above tensor.\rq\rq \cite{Hicks}
\end{quote}
It may thus be useful to discuss another example of torsion.

The simplest non-trivial case of torsion in two-dimensional
manifolds with a positive definite metric is the space of
constant negative curvature viewed as Poincar\'e's upper
half-plane.
Here, the torsion is constant everywhere and homogeneous.

The metric of the space is given by
\begin{equation}
ds^{2} \, = \, \frac{1}{y^{2}} \left( dx^{2} \, + dy^{2} \right) \, , \qquad y \, > \, 0 \, .
\end{equation}
The geodesics appear as semi-circles that intersect the $x$-axis orthogonally.
The upper $y$-axis and all its analogs, $ x = constant, \ y > 0 $,
are special cases of the semi-circles and the geodesics.
In the hyperbolic geometry, they are known as a pencil of parallels
that all intersect in a point at $ y = \infty $.
All this can be derived from the metric.

The lines $ y = constant $ are, however, not of this type.
They are a special case of \lq\lq horocycles\rq\rq\ that appear
in Poincar\'e's map as full circles that touch the $x$-axis
or the point $ y = \infty $.
In the latter case, these horocycles are the lines $ y = constant $,
orthogonal to the pencil of parallel rays $ x = constant $.

We now introduce the frame $ \{ \mbox{\bf e}_{1} , \mbox{\bf e}_{2} \} $ by
\begin{equation}
\mbox{\bf e}_{1} \, = \, y \, \frac{\partial \ }{\partial x} \, , \qquad
\mbox{\bf e}_{2} \, = \, y \, \frac{\partial \ }{\partial y} \, ,
\end{equation}
or the corresponding differential forms
\begin{equation}
{\mbox {\boldmath $\omega$}}_{1} \, = \, \frac{1}{y} \, dx \, , \qquad
{\mbox {\boldmath $\omega$}}_{2} \, = \, \frac{1}{y} \, dy \, , \qquad
ds^{2} \, = \,
({\mbox {\boldmath $\omega$}}_{1})^{2} \, + \,
({\mbox {\boldmath $\omega$}}_{2})^{2} \, .
\end{equation}
(For convenience, we have given the ${\mbox {\boldmath $\omega$}}$'s lower indices.)
[{\bf See Figure~8.}]
\vskip 0.5in
$$
\hskip 0.5 in
{\beginpicture
    \setcoordinatesystem units <1.0in,1.0in> point at 0 0

    \arrow <12pt> [.1,.4] from  -0.5    0.0   to   4.0   0.0  
    \arrow <12pt> [.1,.4] from   0.0   -0.5   to   0.0   5.0  

    \put {\large y} at -0.2    4.75  
    \put {\large x} at  3.75  -0.20  

    \arrow <12pt> [.1,.4] from   0.0  1.0  to  0.0   2.0  
    \arrow <12pt> [.1,.4] from   2.0  1.0  to  2.0   2.0  
    \arrow <12pt> [.1,.4] from   0.0  2.0  to  0.0   3.5  
    \arrow <12pt> [.1,.4] from   0.0  2.0  to  2.0   2.0  
    \arrow <12pt> [.1,.4] from   2.0  2.0  to  2.0   3.5  
    \arrow <12pt> [.1,.4] from   0.0  1.0  to  1.0   1.0  
    \arrow <12pt> [.1,.4] from   2.0  1.0  to  3.0   1.0  
    \arrow <12pt> [.1,.4] from   2.0  2.0  to  4.0   2.0  


    \setquadratic    \setlinear

    \put {$\bullet$}  at 0 0  

    \put {$\bullet$}  at 0.0 1.0  
    \put {$\bullet$}  at 0.0 2.0  
    \put {$\bullet$}  at 2.0 2.0  
    \put {$\bullet$}  at 1.0 1.0  
    \put {$\bullet$}  at 2.0 1.0  

    \put {{\large e}$_1$(1)} at 2.5 0.85
    \put {{\large e}$_1$(2)} at 2.8 2.1
    \put {{\large e}$_2$(1)} at 2.2 1.5
    \put {{\large e}$_2$(2)} at 2.2 2.7


\endpicture}
\vspace{0.5in}
$$
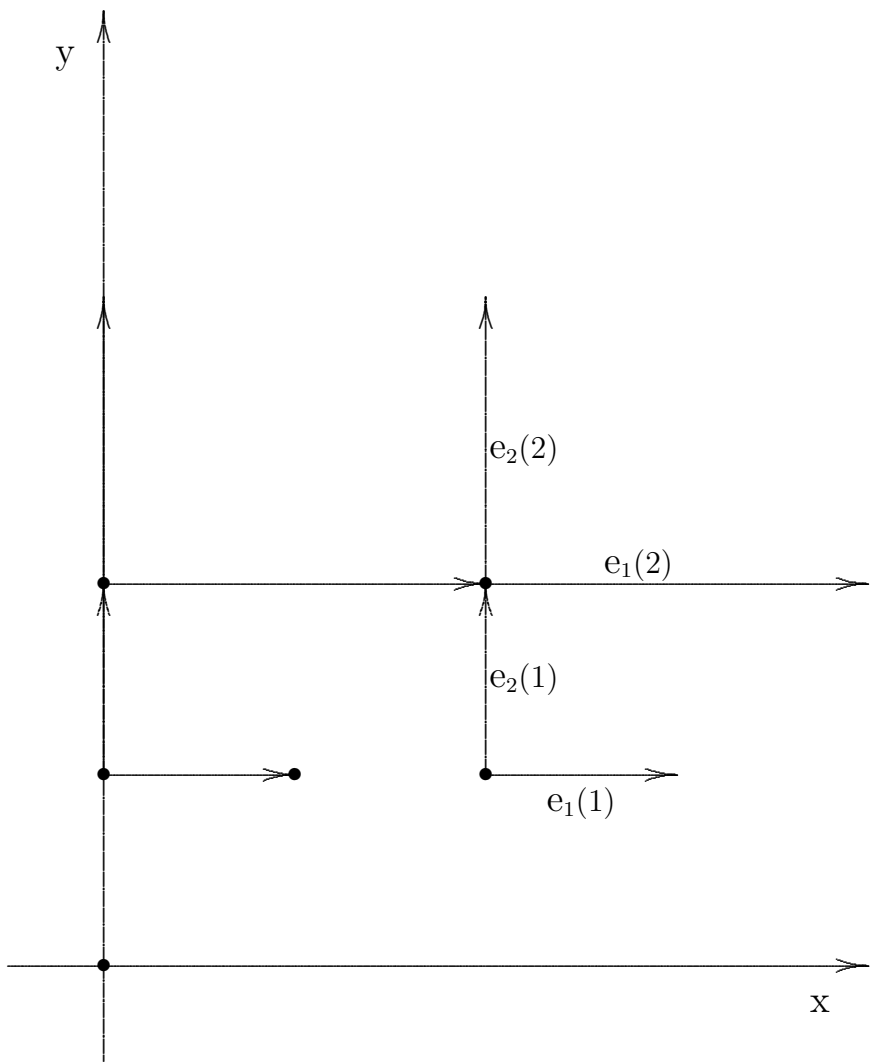
{\centerline {\bf Figure 8: The Notion of Torsion (1)}} \newline

\vfill
\eject
We then have
\begin{equation}
{\bf d}{\mbox {\boldmath $\omega$}}_{1} \, = \,
- \, \frac{1}{y^{2}} \, dy \wedge dx \, = \,
{\mbox {\boldmath $\omega$}}_{1} \wedge {\mbox {\boldmath $\omega$}}_{2} \, = \,
{\mbox {\boldmath $\omega$}}_{12} \wedge {\mbox {\boldmath $\omega$}}_{2} \, ,
\end{equation}
\begin{equation}
{\bf d}{\mbox {\boldmath $\omega$}}_{2} \, = \, 0 \, = \,
{\mbox {\boldmath $\omega$}}_{21} \wedge {\mbox {\boldmath $\omega$}}_{1} \, .
\end{equation}
It follows that the Levi-Civita connection form is given by
\begin{equation}
{\mbox {\boldmath $\omega$}}_{12} \, = \, {\mbox {\boldmath $\omega$}}_{1} \, , \qquad
{\mbox {\boldmath $\omega$}}_{12} \, = \, g_{12\mu} \, {\mbox {\boldmath $\omega$}}_{\mu} \, , 
\end{equation}
which gives
\begin{equation}
g_{121} \, = \, 1 \, , \qquad g_{122} \, = \, 0 \, .
\end{equation}
The connection field is constant.

We obtain, as the only non-zero component of the curvature form,
\begin{equation}
{\mbox {\boldmath $\Omega$}}_{12} \, = \,
- \, {\bf d}{\mbox {\boldmath $\omega$}}_{12} \, = \,
- \, {\bf d}{\mbox {\boldmath $\omega$}}_{1}  \, = \,
- \, {\mbox {\boldmath $\omega$}}_{1} \wedge {\mbox {\boldmath $\omega$}}_{2} \, ,
\end{equation}
which tells us that the curvature is minus one ($-1$).

The torsion of our frame appears if we try to create a
rectangle through parallel displacements.
[{\bf See Figure~9.}]
Let us start at (0,1), moving up to (0,2), moving a
distance one unit right to (2,2), moving down to (2,1),
and then moving a distance one unit left to (1,1).
We are left with the gap from (0,1) to (1,1).
\vskip 0.25in
$$
\hskip 0.5 in
{\beginpicture
    \setcoordinatesystem units <1.0in,1.0in> point at 0 0

    \arrow <12pt> [.1,.4] from  -0.5    0.0   to   4.0   0.0  
    \arrow <12pt> [.1,.4] from   0.0   -0.5   to   0.0   3.0  

    \arrow <12pt> [.1,.4] from   0.0  1.0  to  0.0   2.0  %
    \arrow <12pt> [.1,.4] from   0.0  2.0  to  2.0   2.0  %
    \arrow <12pt> [.1,.4] from   2.0  2.0  to  2.0   1.0  %
    \arrow <12pt> [.1,.4] from   2.0  1.0  to  1.0   1.0  %


    \setquadratic    \setlinear

    \put {$\bullet$}  at 0 0  

    \put {$\bullet$}  at 0.0 1.0  
    \put {$\bullet$}  at 0.0 2.0  
    \put {$\bullet$}  at 2.0 2.0  
    \put {$\bullet$}  at 1.0 1.0  
    \put {$\bullet$}  at 2.0 1.0  

    \put {(0,1)} at -0.2 1.0
    \put {(0,2)} at -0.2 2.0
    \put {(2,2)} at  2.2 2.0
    \put {(2,1)} at  2.2 1.0
    \put {(1,1)} at  1.0 0.85

    \put {\large y} at -0.2    2.75
    \put {\large x} at  3.75  -0.2

\endpicture}
\vspace{0.5in}
$$
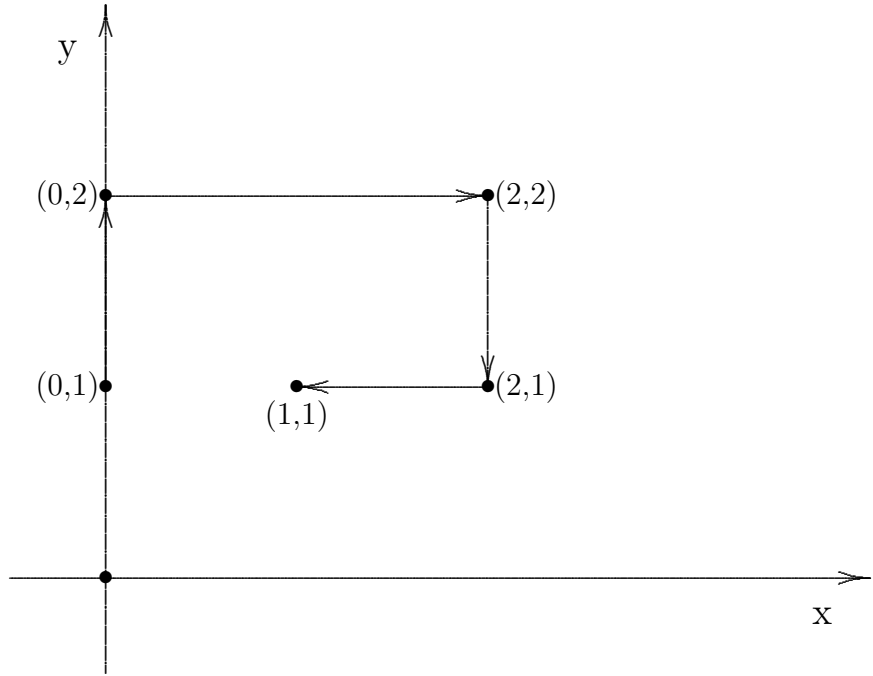
{\centerline {\bf Figure 9: The Notion of Torsion (2)}} \newline


In this example, we study torsion on a manifold with curvature
under the Levi-Civita connection without torsion.
Our frame that gives teleparallelism with a flat (zero) connection
and torsion gives rise to a gap in the parallelogram.
The analog of this gap for the Levi-Civita connection
is {\it not} the phenomenon of curvature but rather the shape
of the geodesics and the parallel propagation of vectors
one-way (not around closed loops).

Clarifying that through geometry, we look at two points
$(x_{1},y_{1})$ and $(x_{2},y_{1})$ equidistant from the $x$-axis.
[{\bf See Figure~10.}]
The circle through these two points that meets the $x$-axis
orthogonally is the Levi-Civita geodesic.
The horocycle $ y_{1} = constant $ is the auto-parallel
connection for the teleparallelism.
The horocycle is not a geodesic; it is a circle with curvature
$ 1 / y_{1} $ and infinite radius.
Parallel transfer of a vector, {\bf V},
initially at $(x_{1},y_{1})$ and pointing in the opposite of the $y$-direction,
changes its orientation by the angle $ - ( x_{2} - x_{1} ) / y_{1} $.
That results from the Levi-Civita connection.
In teleparallelism, there is no change: ${\bf V} = \overline{\bf V}$.
\vskip 0.25in
$$
\hskip 0.0 in
{\beginpicture
    \setcoordinatesystem units <1.0in,1.0in> point at 0 0

    \arrow <12pt> [.1,.4] from  -0.3    0.0   to   4.8   0.0  
    \arrow <12pt> [.1,.4] from   0.0   -0.5   to   0.0   3.0  

    \put {\large y} at -0.2    2.75
    \put {\large x} at  4.7  -0.15

    \circulararc 180 degrees from 4.50 0.0  center at  2.50 0.0



\setsolid
    \plot  1.175  1.50    3.82  1.50  / %
\setdashes
    \plot  0.00   1.50    1.175  1.50  / %
    \plot  1.175  0.00    1.175  1.50  / %
    \plot  3.82   1.50    3.82   0.00  / %

\setdashes <12pt>
    \arrow <12pt> [.1,.4] from   1.175  1.50  to  3.82   1.50  

    \arrow <12pt> [.1,.4] from   1.175  1.50  to  3.35   1.50  
    \arrow <12pt> [.1,.4] from   1.175  1.50  to  3.00   1.50  
    \arrow <12pt> [.1,.4] from   1.175  1.50  to  2.70   1.50  
    \arrow <12pt> [.1,.4] from   1.175  1.50  to  2.35   1.50  
    \arrow <12pt> [.1,.4] from   1.175  1.50  to  2.00   1.50  
    \arrow <12pt> [.1,.4] from   1.175  1.50  to  1.70   1.50  

\setsolid

    \arrow <12pt> [.1,.4] from   1.175  1.50  to  1.175  0.50  
    \arrow <12pt> [.1,.4] from   3.82   1.50  to  3.82   0.50  
    \arrow <12pt> [.1,.4] from   3.82   1.50  to  3.20   0.70  

    \put {$\bullet$}  at 0 0  

    \put {$\bullet$}  at 0.00   1.50  %
    \put {$\bullet$}  at 1.175  0.00  %
    \put {$\bullet$}  at 1.175  1.50  %
    \put {$\bullet$}  at 3.82   0.00  %
    \put {$\bullet$}  at 3.82   1.50  %

    \put {\large y$ _{1} $} at -0.2    1.50
    \put {\large x$ _{1} $} at  1.175 -0.15
    \put {\large x$ _{2} $} at  3.82  -0.15

    \put {\large V}     at  1.28  1.0  %
    \put {$\overline{\bf \large V}$}     at  3.92  1.0  %
    \put {\large V$'$}  at  3.30  1.0  %

    \put {horocycle} at 2.30 1.350  %
    \put {teleparallel transport} at 2.50 1.650  %
    \put {geodesic}     at 2.450 2.10  %
    \put {Levi-Civita}  at 2.450 2.250  %

\endpicture}
\vspace{0.5in}
$$
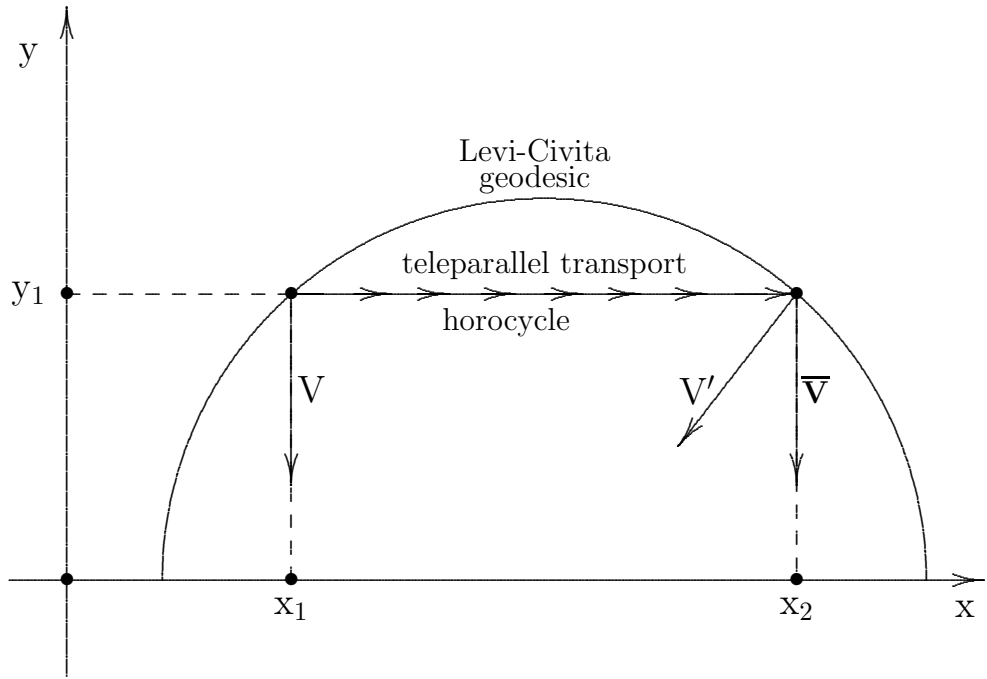
{\centerline {\bf Figure 10: The Notion of Torsion (3)}} \newline
\smallskip
\noindent
When the vector {\bf V} is parallel transported to $(x_2,y_1)$ along the
Levi-Civita geodesic it becomes the vector {\bf V$^\prime$}.
When the vector {\bf V} is teleparallel transported to $(x_2,y_1)$ along the
horocycle $y_1 = constant$ it then becomes the vector $\overline{\bf V}$.

\vfill
\eject
We get the Gauss-Riemann curvature from the Levi-Civita connection
if we consider the rectangle formed by two parallels to the
$y$-axis and their orthogonal horocycles.
[{\bf See Figure~11.}]
Parallel to the $y$-axis the vector {\bf V} does not change
its angle; but along the horocycle $ y_{2} = constant $,
it changes its angle by $ ( x_{2} - x_{1} ) / y_{2} $
which is less for $ y_{2} > y_{1} $ than the change in the
opposite direction along $ y_{1} = constant $.
The overall change in angle is
\begin{equation}
\Delta \phi \, = \, \left( x_{2} - x_{1} \right) \left( \frac{1}{y_{1}} - \frac{1}{y_{2}} \right)
       \, = \, - \, \int \int \frac{dx \wedge dy}{y^{2}} \, .
\end{equation}

The change of angle by parallel transferring a vector {\bf V}
in the positive sense divided by the area gives the curvature
as a constant equal to minus one.

{\bf We see that the curvature is a higher order effect that does
not appear at the level of the torsion.}
It appears as the difference of two parallel transfers.
Although we are dealing with a curved manifold,
all is in perfect analogy to the flat Minkowski plane.

$$
\hskip 0.3 in
{\beginpicture
    \setcoordinatesystem units <1.0in,1.0in> point at 0 0

    \arrow <12pt> [.1,.4] from  -0.3    0.0   to   4.8   0.0  
    \arrow <12pt> [.1,.4] from   0.0   -0.5   to   0.0   4.0  

    \put {\large y} at -0.2    3.75
    \put {\large x} at  4.7  -0.15


\setdashes
    \plot  0.00   1.50    3.82   1.50  / %
    \plot  0.00   3.00    3.82   3.00  / %
    \plot  1.175  0.00    1.175  3.00  / %
    \plot  3.82   3.00    3.82   0.00  / %
\setsolid

    \arrow <12pt> [.1,.4] from   1.175  1.50  to  1.175  0.50  
    \arrow <12pt> [.1,.4] from   3.82   1.50  to  3.20   0.70  
    \arrow <12pt> [.1,.4] from   3.82   3.00  to  3.20   2.20  
    \arrow <12pt> [.1,.4] from   1.175  3.00  to  1.75   2.15  
    \arrow <12pt> [.1,.4] from   1.175  1.50  to  1.75   0.65  

    \arrow <12pt> [.1,.4] from   2.40   1.50  to  2.60   1.50  
    \arrow <12pt> [.1,.4] from   2.60   3.00  to  2.40   3.00  
    \arrow <12pt> [.1,.4] from   3.82   2.20  to  3.82   2.40  
    \arrow <12pt> [.1,.4] from   1.175  2.30  to  1.175  2.10  

    \put {$\bullet$}  at 0 0  

    \put {$\bullet$}  at 0.00   1.50  %
    \put {$\bullet$}  at 0.00   3.00  %
    \put {$\bullet$}  at 1.175  0.00  %
    \put {$\bullet$}  at 1.175  1.50  %
    \put {$\bullet$}  at 1.175  3.00  %
    \put {$\bullet$}  at 3.82   0.00  %
    \put {$\bullet$}  at 3.82   1.50  %
    \put {$\bullet$}  at 3.82   3.00  %

    \put {\large y$ _{1} $} at -0.2    1.50
    \put {\large y$ _{2} $} at -0.2    3.00
    \put {\large x$ _{1} $} at  1.175 -0.15
    \put {\large x$ _{2} $} at  3.82  -0.15

    \put {\large V}     at  1.07  1.0  
    \put {\large V$'$}  at  3.30  1.0  

\endpicture}
\vspace{0.5in}
$$
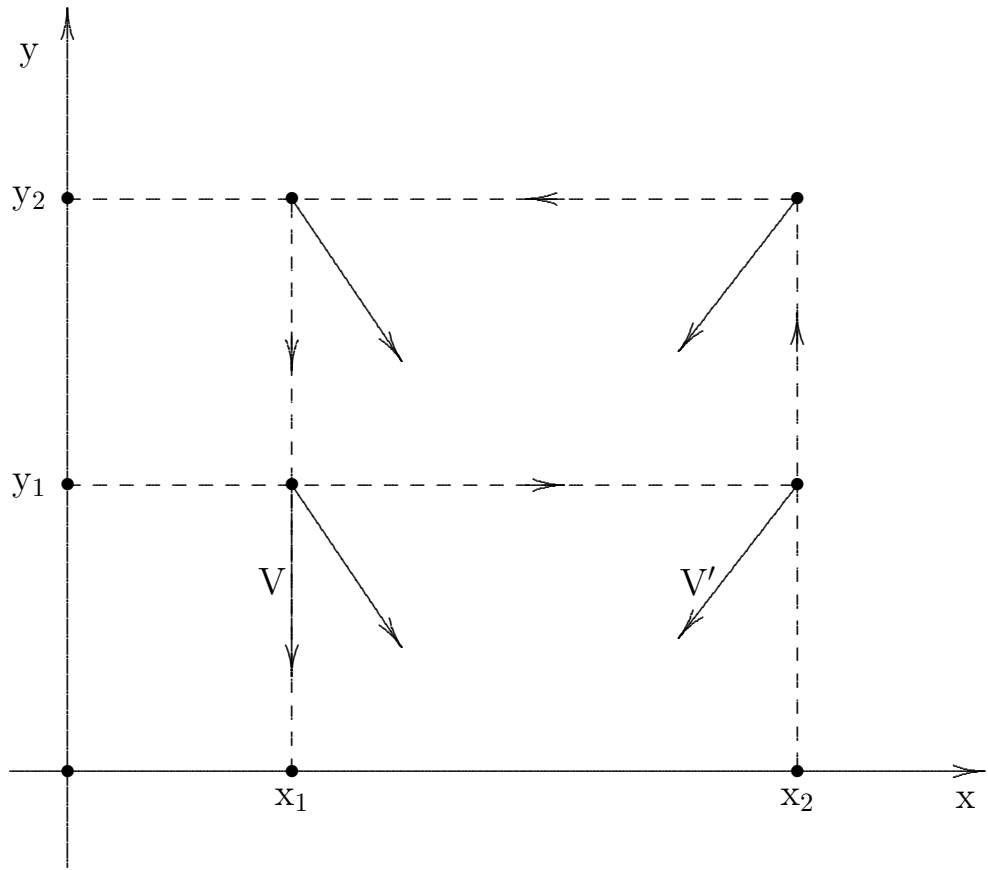
{\centerline {\bf Figure 11: The Notion of Torsion (4)}} \newline


\section{Discussion}
\setcounter{equation}{0}

We now return to the question:
\lq\lq What is the physical meaning of Einstein's first
principle of equivalence?\rq\rq\

In his monograph \lq\lq Relativity, The General Theory\rq\rq\cite{Synge},
John Synge confesses in his introduction
\begin{quote}
\lq\lq \dots I have never been able to understand this Principle.\rq\rq
\end{quote}
and goes on to write:
\begin{quote}
\lq\lq Does it mean that the effects of a gravitational field
are indistinguishable from the effects of an observer's acceleration?
If so, it is false.
In Einstein's theory,
either there is a gravitational field or there is none,
according as the Riemann tensor does or does not vanish.
This is an absolute property;
it has nothing to do with any observer's worldline.
Space-time is either flat or curved, and in several places
of the book I have been at considerable pains to separate
truly gravitational effects due to curvature of space-time
from those due to curvature of the observer's worldline
(in most ordinary cases the latter predominate).
The Principle of Equivalence performed the essential office
of midwife at the birth of general relativity,
but, as Einstein remarked, the infant would never have got
beyond its long-clothes had it not been for Minkowski's concept.
I suggest that the midwife be now buried with appropriate honours
and the facts of absolute space-time faced.\rq\rq
\end{quote}
Are we beginning a chapter of \lq\lq forensic physics\rq\rq\
if we investigate the corpse of a principle?
It is easy to agree with Synge.
If one admits a metric, one also buys into the Levi-Civita
connection and its Riemann tensor and this connection makes
itself felt.
But that was not all.
From the beginning of relativity, already in the special theory,
there was the question of the reference body.
A reference body can be defined mathematically through a section
of the frame bundle of the spacetime manifold, with
Fermi's construction being an approximation for physics in some cases.
So, it is the mathematical reference body, the generalization
of the constant parallel orthonormal 4-vectors in Minkowski
space-time that now show teleparallelism and torsion.
Mathematically, we can formulate our principle as
\begin{equation}
{\bf \Theta}^{\mu} \, = \,
{\bf d}{\mbox {\boldmath $\omega$}}^{\mu} \, = \,
{\mbox {\boldmath $\omega$}}^{\mu}{}_{\nu} \wedge {\mbox {\boldmath $\omega$}}^{\nu} \, .
\end{equation}
Its meaning is that we enlarge the set of admissible frames
for the description of physical phenomena in spacetime.

Einstein tried to understand gravitation through acceleration
and that made it necessary to introduce accelerated frames,
that is, frames with torsion.

Although Gregorio Ricci-Curbastro had already introduced
frames into manifolds in 1895,
their use appeared optional until the advent of Paul Dirac's
equation for spin $1/2$ particles.
Most physicists approached the interpretation of general
relativity from the particle point of view.
But fields are more important than particles---at least quantum fields.
Fermi was an exception.
In his study of Fermi transport, he worried about the
description of an electromagnetic field in an accelerated
reference system.
How would the energy density be affected by the acceleration?
This raises the general question of how fields are affected by torsion.
The prime example for this is the Unruh effect \cite{Unruh}
where the field is the vacuum.
It is clear that a mathematical section of the frame bundle
does not give rise to physical effects like energy densities, etcetera.
But as soon as one attaches physical objects to the frames,
one has reference bodies and for the Unruh effect \cite{Wald},
what one calls \lq\lq particle detectors.\rq\rq\
But, we have already seen that where light is emitted by a source
at rest and absorbed by a receiver also at rest,
we get redshifts in accelerated systems as measured by Pound and Rebka.

\section{Conclusion}
\setcounter{equation}{0}

In 1907 Einstein did not show the equivalence of acceleration
and gravitation described by spacetime curvature.
He did not show either the equivalence of geodesics and non-geodesics
or the equivalence of rotating and non-rotating systems.
What he did, we now can see more clearly, was the introduction
of accelerated reference systems exhibiting torsion through
distant parallelism.
There were physical consequences that needed to be checked
for these systems, like the constancy of the speed of light
independent of acceleration, no influence of acceleration on
the rate of clocks and the length of standards.
As far as these assumptions have been tested,
they appear to be in order.

In 1911 Einstein formulated an equivalence principle that
involved relative acceleration in an attempt to introduce
ideas of Ernst Mach into his theory \cite{Einstein1911}.
This did not prove to be a happy idea since this notion makes
mathematical sense only for bodies having the same 4-velocity
and from a physical point of view accelerations are absolute.
These ideas gave the theory its name.

However, as John Stachel \cite{Stachel} pointed out, 
it was the older idea of 1907 that guided him through
Ehrenfest's paradox of the rotating disc to Riemannian geometry.
We can now see that going from Levi-Civita's connection in
Minkowski's spacetime to teleparallelism
opens the door to going back from torsion
to a Levi-Civita connection with curvature.
Einstein's apple was, like Newton's apple, a seminal thought
of great penetrating power.
It has been claimed that the story of Newton's apple
was apocryphal; but a book containing the recollections of
William Stukeley \cite{Stukeley} appears to confirm it.
Stukeley, a doctor, from Lincolnshire like Newton,
became a close friend to Newton in Sir Isaac's last years.
Stukeley describes the Summer evening when Newton,
then in his eighties, recalled his thoughts from sixty years before:
\begin{quote}
\lq\lq After dinner, the weather being warm, we went into the
garden and drank tea, under the shade of some apple trees,
only he and myself.
Amidst other discourse, he told me, he was just in the same situation,
as when formerly, the notion of gravitation came into his mind.
It was occasion'd by the fall of an apple,
as he sat in a contemplative mood.
Why should that apple always descend perpendicularly to the ground,
thought he to himself.
Why should it not go sideways or upwards,
but constantly to the earth's centre?
Assuredly the reason is, that the earth draws it.
There must be a drawing power in matter:
and the sum of the drawing power must be in the earth's centre,
not in any side of the earth.
Therefore does the apple fall perpendicularly, or towards the centre.
If matter thus draws matter, it must be in proportion of its quantity.
Therefore the apple draws the earth, as well as the earth draws the apple.
That there is a power, like that we here call gravity,
which extends itself into the universe.\rq\rq
\end{quote}
What Newton had done was to look at gravity from a new frame
whose origin was in the center of the Earth.



\vfill
\end{document}